\documentclass[twocolumn,aps,showpacs,floatfix,prc]{revtex4}
\usepackage[dvips]{epsfig}
\begin{document}
\title{Compact star deformation and universal relationship for magnetized white dwarfs}

\author{Sujan Kumar Roy$^{1*\S}$, Somnath Mukhopadhyay$^{2\S\dagger}$ and D. N. Basu$^{3*\S}$}

\affiliation{$^*$ Variable  Energy  Cyclotron  Centre, 1/AF Bidhan Nagar, Kolkata 700 064, India }
\affiliation{$^{\S}$ Homi Bhabha National Institute, Training School Complex, Anushakti Nagar, Mumbai 400085, India}
\affiliation{ $^\dagger$ National Institute of Technology, Tiruchirapalli - 620015, Tamil Nadu, India}

\email[E-mail 1: ]{sujan.kr@vecc.gov.in}
\email[E-mail 2: ]{smpapan7@gmail.com}
\email[E-mail 3: ]{dnb@vecc.gov.in}
\date{\today }

\begin{abstract}

    Recently super-Chandrasekhar mass limit has been derived theoretically in presence of strong magnetic field to complement experimental observations. In the framework of Newtonian physics, we have studied the equilibrium configurations of such magnetized white dwarfs by using the relativistic Thomas-Fermi equation of state for magnetized white-dwarfs. Hartle formalism, for slowly rotating stars, has been employed to obtain the equations of equilibrium. Various physical quantities of uniformly rotating and non-rotating white dwarfs have been calculated within this formalism. Consequently, the universality relationship between the moment of inertia(I), rotational love number($\lambda$) and spin induced quadrupole moment(Q), namely the I-Love-Q relationship, has been investigated for such magnetized white dwarfs. The relationship between I, eccentricity and Q i.e. I-eccentricity-Q relationship has also been derived. Further, we have found that, the I-eccentricity-Q relationship is more universal in comparison to I-Love-Q relationship.
          
\vspace{0.2cm}    

\noindent
\end{abstract}


\maketitle

\noindent
\section{Introduction}
\label{Section 1} 
  Black holes are said as bald because their multipole moments can be expressed in terms of their mass, charge and spin angular momentum. There are numbers of studies carried out recently on slowly rotating neutron stars, quark stars and dark stars, which show the equation of state independent nature in the relations between certain multipole moments of these stars \cite{Yagi2014, Yagi2013, Yagi2013b, Yagi2014b, Reina2017, Maselli2017, Yip2017, Sham2015, Maselli2013, Pappas2014}. Binary neutron stars are one of the most promising gravitational wave (GW) sources \cite{Sathya2009, Abadie2010} for Advanced LIGO, Advanced VIRGO, KAGRA, Einstein telescope. Neutron star binaries can be used to extract information about the equation of state (EoS) by detecting the GWs emitted in the late inspiral during which neutron stars are tidally deformed. Therefore, GWs emitted by neutron star binaries in the late inspiral must incorporate corrections induced by the neutron star internal structure, thereby providing information about the EoS \cite{Hinderer2008, Damour2009, Binnington2009}. Unlike black holes, exterior gravitational fields of neutron stars are not only determined by their mass, radius, spin angular momentum but also by their higher multipole moments. The extraction of the higher multipole moments from observations might be erroneous if EoS dependent description is neglected. But, degeneracies between the neutron star quadrupole moment and the spin prevent detections from separately measuring these quantities. Again, degeneracies between the effect of the neutron star EoS and corrections due to modified gravity on observables prevent serious tests of general relativity which are internal structure independent. 
  
  The I-Q relationship breaks down as the rotation increases to the mass-shedding limit and the deviation may be as high as 40\% for neutron stars and 75\% for strange stars \cite{Doneva2014}. Considering the effect of strong magnetic field on slowly rotating neutron stars it has been found that, for the case of a realistic magnetic field configuration, the relation again depends significantly on the EoS, losing its universality \cite{Haskell2014}. Therefore, the use of the universality relationships in the subject of GW physics is limited by different conditions like rotation, magnetic field etc.
  
  White dwarfs are another important type of compact stars. As the most common end product of evolution scenario, white dwarfs account for almost 97\% of all evolved stars. Therefore, the properties and distribution of white dwarfs contain plenty of information. Consequently, the detections of white dwarfs and study of their properties are of paramount importance. From GW physics point of view, white dwarfs in binaries are potential source of the powerful gravitational waves for LIGO/VIRGO type interferometers \cite{Lipunov2017}. Therefore, in the similar fashion of neutron stars the study of relationship between different parameters of white dwarfs under different conditions has attracted certain attention in the literature \cite{BoshK2017,BoshK2018}. It has already been shown in Ref.-\cite{BoshK2017} that, the universality relationships hold good between the moment of inertia, tidal love number and spin induced quadrupole moment for the case of cold white dwarfs as well. But as the effect of temperature is incorporated into the EoS it turns out that, the universality breaks down as the temperature approaches millions of Kelvin \cite{BoshK2018}. As, whether these universality relationships hold in presence of magnetic field is yet to be investigated, in this work we have studied the effects of magnetic field on these relationships. The effects of the magnetic field have been incorporated to obtain the EoS for the Helium, Carbon, Oxygen, Iron white dwarfs and consequently, in the framework of Newtonian physics equilibrium configurations of such magnetized white dwarfs have been obtained. Finally, all the physically relevant parameters and multipole moments have been calculated to check out the validity of the universal relationships between different multipole moments. 
  
\noindent
\section{Formalism and Equations of equilibrium} 
\label{Section 2}

  The formalism we have used in this work, is that devised by Hartle \cite{Hartle1967} for uniformly rotating compact stars adopted in the framework of Newtonian gravity. Therefore, the basic formalism to determine the equilibrium configurations would assume the slow rotation approximation within the framework of Newtonian physics and thus the rotation in such slowly, uniformly rotating stars will be treated as perturbation to the non-rotating spherical stars. As white dwarfs can have low compactness compared to neutron stars, they can be well treated within the framework of Newtonian physics as the general relativistic corrections for various physical quantities turn out to be very small \cite{BoshK2014,BoshK2017}.
  
  It is worth mentioning the fact that, this particular approach becomes advantageous in the sense that, from physical and geometric point of view it is very much intuitive because of the use of coordinates make direct correspondence to the configuration and dynamics of rotating bodies in equilibrium. The relativistic Hartle formalism, when applied in the context of Newtonian gravity, the equilibrium configurations can be described through ordinary differential equations. Moreover, these equations i.e. equations of equilibrium, can be solved with required physical conditions to extract all the relevant physical quantities such as the mass $M$, the equatorial radius $r_e$, the polar radius $r_p$, the moment of inertia $I$, angular momentum $J$, ellipticity $\epsilon$ and the tidal love number $\lambda$ of the star.
  
\subsection{Slowly and uniformly rotating stars in Newtonian gravity}
\label{Section 2A}

    When a star is said to be rotating slowly and uniformly, it means that all the particles within the star must move with a speed much lower than the speed of light. This can be simply represented by,
    
\begin{equation}
   \Omega^2<<\frac{GM}{R^3},
\label{eqn1}  
\end{equation}
where $\Omega$ is the angular velocity of the star, $G$ is the universal gravitational constant, $M$ is the mass and $R$ is the radius of the  star. Therefore, Eq.(\ref{eqn1}) is the condition under which we can treat the rotation as a small perturbation to the already known non-rotating spherical configuration having axial and reflection symmetry.

  For a rotating configuration, each of the particle of the star in spherical polar coordinate ($r$,$\theta$,$\phi$) will acquire a centrifugal force given by,
   \begin{equation}
     \left(\frac{d^2\vec{r}}{dt^2}\right)_{C}=-\hat{r} r\Omega^2 \sin^2\theta-\hat{\theta} r\Omega^2 \sin\theta \cos\theta ,
   \label{eqn2}
   \end{equation}
  which incurs no $\phi$ dependent force, thus retaining the axial symmetry.  Moreover, as the force terms depend only on $\Omega^2$, the configuration will be symmetric under the reversal of the rotation axis as well. Hence, the rotating configuration will have reflection symmetry as well.
  
  Hence, the equation of gravitational potential $\Phi(r,\theta)$ becomes,
   \begin{equation}
     \nabla^2 \Phi(r,\theta)=4\pi G\rho(r,\theta),
   \label{eqn3}
   \end{equation}
   where $\rho(r,\theta)$ is the mass density of the star. The $\theta$ dependence in the equation is the consequence of the rotation of the star.
   
   The EoS for the star is assumed to be of the form $p=p(\rho)$, i.e. for the cold matter. Then the equation of the hydrostatic equilibrium turns out to be,
    \begin{eqnarray}
   && -\frac{\nabla p(r,\theta)}{\rho(r,\theta)}-\nabla \Phi(r,\theta)= \left(\frac{d^2\vec{r}}{dt^2}\right)_C \nonumber\\
   && \Rightarrow \int \frac{dp(r,\theta)}{\rho(r, \theta)}+\Phi(r,\theta)= \frac{1}{2}\Omega^2 r^2 \sin^2\theta+Constant. \nonumber\\
    \label{eqn4}
    \end{eqnarray}
    The rotating configurations of stars in equilibrium can be found by solving Eqs.-(\ref{eqn3}) \& (\ref{eqn4}) by assuming that, all the points on a spherical surface of a particular density in non-rotating configuration now lie on a deformed surface of the same density in the rotating configuration. Assuming slow rotation, the introduction of the Hartle coordinate for the rotating deformed star gives,
    \begin{equation}
     r=R+\xi; \; \; \; \Theta=\theta,
    \label{eqn5}
    \end{equation}
    where $R$, $\Theta$ represent a point in the non-rotating configuration and the same point is represented by $r$,$\theta$ in the rotating configuration. As the rotating configuration of the star retains axial and reflection symmetries there would be no $\phi$ dependence in the equations and for the same reason the perturbation term $\xi$ can contain only $\Omega^2$ terms. Hence, the perturbations in radius, mass, potential, moment of inertia due to rotation can be represented in terms of $\Omega^2$ powers. Throughout this work perturbation terms only up to first power in $\Omega^2$ will be considered for calculation purpose. Hence, Eq.(\ref{eqn5}) can be represented as below,
    \begin{equation}
     r(R,\Theta)=R+\xi(R,\Theta)+O(\Omega^4); \; \; \; \Theta=\theta,
    \label{eqn6}
    \end{equation}
    where the term $\xi(R,\Theta) \sim \Omega^2$ and for the slow rotation approximation to be valid the condition need to be fulfilled is,
    \begin{equation}
     \frac{\xi(R,\Theta)}{R}<<1, \; \; \forall R \;.
    \label{eqn7}
    \end{equation}
      
    Corresponding to all these, the density profile and consequently the EoS for the rotating configuration of the star can be represented as follows,
    \begin{eqnarray}
     && \rho(r,\theta)=\rho(R,\Theta)=\rho(R)=\rho^{(0)}(R), \nonumber\\
     && p(\rho)=p(R,\Theta)=p(R)=p^{(0)}(R),
    \label{eqn8}
    \end{eqnarray}
    where $\rho^{(0)}(R)$ and $p^{(0)}(R)$ are the corresponding quantities in the non-rotating configuration.\\ 
     
\subsection{Equations of Equilibrium}
\label{Section 2B}
    Now, the equations of equilibrium for the rotating configurations can be found by transforming the Eqs.-(\ref{eqn3}) \& (\ref{eqn4}) in $R, \Theta$ coordinate and decomposing deformation $\xi$, potential $\Phi$ as,
    \begin{eqnarray}
     && \xi(R,\Theta)=\sum_{l}\xi_l(R)P_l(\cos\Theta), \nonumber\\
     && \Phi(R,\Theta)=\Phi^{(0)}(R)+\Phi^{(2)}(R,\Theta)+O(\Omega^4), \nonumber\\
     && \Phi^{(2)}(R,\Theta)=\sum_{l}\Phi_l^{(2)}(R) P_l(\cos\Theta).
    \label{eqn9}
    \end{eqnarray}
    $P_l(\cos\Theta)$ is the Legendre polynomial of the order $l$ and $\Phi^{(0)}(R)$ is the potential in the non-rotating configuration, whereas, the $\Phi^{(2)}(R,\Theta)$ term in rotating configuration represents the perturbation $\sim	\Omega^2$. Ultimately, these yields
    \begin{eqnarray}
     \nabla^2\Phi^{(0)}(R)=4\pi G\rho^{(0)}(R), \;\;\;\; (\; \Omega^0 \;),
    \label{eqn10}
    \end{eqnarray}
     \begin{eqnarray}
     \xi_0(R)\frac{d}{dR}\nabla^2\Phi^{(0)}(R)+\nabla^2\Phi_0^{(2)}(R)=0, \nonumber\\
     \;\;\;\;\;\;\;\;\;\;\;\;\;\;\;\;\;\;\;\;\;\;\;\;\;\;\;\;\;\;\;\;\;\; (\Omega^2 \;\; \& \;\; l=0),
    \label{eqn11}
    \end{eqnarray}
    \begin{eqnarray}
     && \xi_2(R)\frac{d}{dR}\nabla^2\Phi^{(0)}+\nabla^2\Phi_2^{(2)}(R)-\frac{6}{R^2}\Phi_2^{(2)}(R)=0, \nonumber\\
     && \;\;\;\;\;\;\;\;\;\;\;\;\;\;\;\;\;\;\;\;\;\;\;\;\;\;\;\;\;\;\;\;\;\;\;\;\;\;\;\;\;\; (\Omega^2 \;\; \& \;\; l=2),
    \label{eqn12}
    \end{eqnarray}
    
    from Eq.-(\ref{eqn3}) and from Eq.-(\ref{eqn4})
    \begin{equation}
     \int_{0}^{p} \frac{dp^{(0)}(R)}{dR}+\Phi^{(0)}(R)=Constant,\;\;\;\; (\; \Omega^0 \;),
    \label{eqn13}
    \end{equation}
    \begin{eqnarray}
     \xi_0(R)\frac{d\Phi^{(0)}(R)}{dR}+\Phi_0^{(2)}(R)-\frac{1}{3}\Omega^2R^2=0, \nonumber\\
     \;\;\;\;\;\;\;\;\;\;\;\;\;\;\;\;\;\;\;\;\;\;\;\;\;\;\;\;\;\;\;\;\;\; (\Omega^2 \;\; \& \;\; l=0),
    \label{eqn14}
    \end{eqnarray}
    \begin{eqnarray}
     \xi_2(R)\frac{d\Phi^{(0)}(R)}{dR}+\Phi_2^{(2)}(R)+\frac{1}{3}\Omega^2R^2=0, \nonumber\\
     \;\;\;\;\;\;\;\;\;\;\;\;\;\;\;\;\;\;\;\;\;\;\;\;\;\;\;\;\;\;\;\;\;\; (\Omega^2 \;\; \& \;\; l=2).
    \label{eqn15}
    \end{eqnarray}
     As the Eq.-(\ref{eqn4}) contains only $\sin^2\Theta$ term in the right hand side we get equations for $l=$ 0 \& 2 only. The Eqs.(\ref{eqn10}-\ref{eqn15}) altogether describe the equilibrium configuration of the rotating star with $\xi_l$ and $\Phi^{(2)}_l$ $(l=0,2)$ being the unknowns and
     \begin{equation}
      \xi_l=0, \;\;\;\; \Phi^{(2)}_l=0, \;\; for \;\; l>2
     \label{eqn15a}
     \end{equation}
\subsection{Equations of Spherical Background}
\label{Section 2C}
     The equations corresponding to $\Omega^0$ 	i.e. Eqs.-\ref{eqn10} \& \ref{eqn13} together represent the background spherical configuration or the non-rotating configuration on which the rotation is assumed to be a perturbation. Simplification of these equations gives,
     \begin{eqnarray}
    && \frac{dp^{(0)}(R)}{dR}=-\rho^{(0)}(R) \frac{GM^{(0)}(R)}{R^2}, \nonumber\\
    && \frac{dM^{(0)}(R)}{dR}=4\pi R^2\rho^{(0)}(R),      
     \label{eqn16}
     \end{eqnarray}
    which is the non-relativistic limit ($c$$\rightarrow$$\infty$) of the Tolman-Oppenheimer-Volkov \cite{TOV39a,TOV39b} equations. $M^{(0)}(R)$ in these equations, is simply the mass within radius $R$ of the spherical configuration and corresponding moment of inertia $I^{(0)}(R)$ can be found from the equation,
    \begin{equation}
     I^{(0)}(R)=\frac{8\pi}{3}\int_{0}^{R}\rho^{(0)}(R)R^4 dR
    \label{eqn17}
    \end{equation}
    Boundary conditions used are, $\rho^{(0)}(R)\rightarrow \rho_c$ \& $M\rightarrow 0$ while $R\rightarrow 0$ and $p^{(0)} \rightarrow 0$ as $R\rightarrow a$, with $a$ being the radius of the unperturbed star and $\rho_c$ is the central density of the configuration. It is worth mentioning the fact that, the Hartle formalism allows one to determine the physical quantities like mass, radius, moment of inertia etc, for the rotating configuration with the same central density $\rho_c$ as the non-rotating configuration. It also turns out that, the gravitational potential inside the star is connected to the mass in the following way,
    \begin{equation}
     \frac{d\Phi^{(0)}(R)}{dR}=\frac{GM^{(0)}(R)}{R^2}
    \label{eqn18}
    \end{equation}
    
\subsection{Equations of l=0 Deformation}
\label{Section 2D}
    The total mass of the star $M_{tot}(R)$ of the rotating star can be defined as follows,
    \begin{eqnarray}
     && M_{tot}(R)=\int_{V} \rho(r,\theta) r^2 \sin\theta dr d\theta d\phi \nonumber\\
     =&&\int_{V}\rho^{(0)}(R) (R+\xi)^2 \sin\Theta (dR+d\xi)d\Theta d\phi \nonumber\\
     =&&\int_{V}\rho^{(0)}(R)R^2 \sin\Theta dR d\Theta d\phi + \nonumber\\
     &&\int_{V}\rho^{0}(R)R^2\left(\frac{2\xi(R,\Theta)}{R}+\frac{d\xi(R,\Theta)}{dR}\right)\sin\Theta dR d\Theta d\phi. \nonumber\\
    \label{eqn18a}
    \end{eqnarray}
    The first integral in the right hand side of this equation is clearly providing the definition of $M^{(0)}(R)$ as given in Eq.-(\ref{eqn16}). The second integral is defined as $M^{(2)}(R)$ and can be further simplified with the help of the decomposition of $\xi(R,\Theta)$ (Eq.-(\ref{eqn9})), as follows,
    \begin{eqnarray}
     M^{(2)}(R)=&& 4\pi\int_{0}^{R} \rho^{0}(R)R^2\left(\frac{2\xi_0(R)}{R}+\frac{d\xi_0(R)}{dR}\right) dR \nonumber\\
     =&& 4\pi\int_{0}^{R} \left(-\xi_{0}(R)\frac{d\rho^{(0)}(R)}{dR}\right)R^2 dR.
    \label{eqn18b}
    \end{eqnarray}
    Eqs.-\ref{eqn11} \& \ref{eqn14} provide the perturbation terms $\sim \Omega^2$ corresponding to $l=0$, represent the spherical deformation of the star due to the rotation. Simplifying these two equations and incorporating the definition of $M^{(2)}(R)$ we get, 
    \begin{eqnarray}
    -\frac{dp^{*}_0(R)}{dR}+\frac{2}{3}\Omega^2R=\frac{GM^{(2)}(R)}{R^2}, \nonumber\\
     \frac{dM^{(2)}(R)}{dR}=4\pi R^2\rho^{(0)}(R)\frac{d\rho^{(0)}}{dp^{(0)}}p^{*}_0(R),
    \label{eqn19}
    \end{eqnarray}
    where $M^{(2)}(R)$, the extra mass of the rotating star over its spherical background, can be supported with the centrifugal force while the rotating star is having the same central density as its non-rotating background. The new variable $p^{*}_0(R)$ is defined by,
    \begin{equation}
     p^{*}_0(R)=\xi_0(R)\frac{d\Phi^{(0)}(R)}{dR}.
    \label{eqn20}
    \end{equation}
    The equation set Eq.-\ref{eqn19} is solved with the boundary conditions $M^{(2)}(R)\rightarrow 0$ while $R\rightarrow 0$ and $p^{*}_0(R) \rightarrow \frac{1}{3}\Omega^2R^2$ as $R\rightarrow 0$. The perturbation in the potential $\Phi^{(2)}_0(R)$ is then given by, 
    \begin{equation}
     \frac{d\Phi^{(2)}_0(R)}{dR}=\frac{GM^{(2)}(R)}{R^2}
    \label{eqn21}
    \end{equation}
    
\subsection{Equations of l=2 Deformation}
\label{Section 2E}
    Eqs.-\ref{eqn12} \& \ref{eqn15}, the equations provide the perturbation terms $\sim \Omega^2$ corresponding to $l=2$, represent the quadrupolar deformation of the star due to the rotation. Simplification of these equations and introduction of the new variable $\chi$ provide,
    \begin{eqnarray}    
     &&\frac{d\chi(R)}{dR}=-\frac{2GM^{(0)}(R)}{R^2}\Phi^{(2)}_{2}(R)+\frac{8\pi}{3}G\Omega^2\rho^{(0)}(R)R^3, \nonumber\\
     &&\frac{d\Phi^{(2)}_{2}(R)}{dR}=\left(\frac{4\pi R^2\rho^{(0)}(R)}{M^{(0)}(R)}-\frac{2}{R}\right)\Phi^{(2)}_{2}(R) \nonumber\\
     &&+\frac{4\pi}{3M^{(0)}(R)}\Omega^2 \rho^{(0)}(R)R^4-\frac{2\chi(R)}{GM^{(0)}(R)}.
    \label{eqn22}
    \end{eqnarray}
    This set of equations are integrated numerically outwards with necessary conditions which turn out to be, as $R\rightarrow 0$, 
    \begin{equation}
     \Phi^{(2)}_{2}(R) \rightarrow AR^2 \;\; \& \;\; \chi(R)\rightarrow BR^4,
    \label{eqn23}
    \end{equation}
    where $A$ and $B$ are arbitrary constants related through,
    \begin{equation}
     \frac{2\pi G}{3}A\rho_c+B=\frac{2\pi G}{3}\Omega^2 \rho_c.
    \label{eqn24}
    \end{equation}
    The other condition applies is, $\Phi^{(2)}_{2}(R) \rightarrow 0$ ,as $R \rightarrow \infty$. This large $R$ behavior of Eqs.-(\ref{eqn22}) shows, as $R \rightarrow \infty$, 
    \begin{equation}
     \chi(R) \rightarrow K_{1}\frac{GM^{(0)}(a)}{2R^4} \;\; \& \;\; \Phi^{(2)}_{2}(R)\rightarrow \frac{K_{1}}{R^3},
    \label{eqn25}
    \end{equation}
    where the $M^{(0)}(a)$ is the total mass of the unperturbed star and $K_1$ is the arbitrary constant need to be determined from the condition of continuity of $\chi(R)$ and $\Phi^{(2)}_{2}(R)$ at the surface of the star.
    
\section{Calculation of physical quantities} 
\label{Section 3}

    Having determined all the equations of structure, in this section all the required physical quantities for the present work will be summarized.
    
\subsection{Mass and Radius}
\label{Section 3A}
    Following the coordinate transformation in Eq.-(\ref{eqn5}) the radius of the star at surface can be expressed as,
    \begin{equation}
     r(a,\Theta)=a+\xi_0(a)P_0(\cos\Theta)+\xi_2(a)P_2(\cos\Theta),
    \label{eqn26}
    \end{equation}
    where $\Theta=0$ or $\pi$ represent the polar points and the equator is represented by $\Theta=\pi/2$. Hence, the polar radius $r_p$ and the equatorial radius $r_e$ are given by,
    \begin{eqnarray}
     r_p=&& r(a,0)=r(a,\pi)=a+\xi_0(a)+\xi_2(a), \nonumber\\
     r_e=&& r(a,\pi/2)=a+\xi_0(a)-\xi_2(a)/2,
    \label{eqn27}
    \end{eqnarray}
    where $\xi_0(a)$ is found from Eq.-(\ref{eqn20}). As the solution of $\Phi^{(2)}_2(R)$ is found by solving Eq.-(\ref{eqn22}), $\xi_2(a)$ can easily be determined using Eq.-(\ref{eqn15}), i.e.
    \begin{equation}
     \xi_2(a)=-\left(\frac{\Phi^{(2)}_2(R)+\frac{1}{3}\Omega^2R^2}{\frac{d\Phi^{(0)}(R)}{dR}}\right)_{R=a}
    \label{eqn28}
    \end{equation}
    The eccentricity of the rotating spheroid is given through,
    \begin{equation}
     e=\sqrt{1-\left(\frac{r_p}{r_e}\right)^2}.
    \label{eqn29}
    \end{equation}
    The definition of the total mass is given in Eq.-(\ref{eqn18a}). Hence, we can write the total mass of the rotating configuration $M_{tot}$ as,
    \begin{eqnarray}
     && M_{tot}=M^{0}(a)+M^{2}(a) \\
     && =\int_{0}^{a}4\pi R^2 \rho^{(0)}(R)dR + \int_{0}^{a}4\pi R^2 \rho^{(0)}(R) \frac{d\rho^{(0)}}{dp^{(0)}} p^{(*)}_0(R) dR. \nonumber
    \label{eqn30}
    \end{eqnarray}
\subsection{Total Moment of Inertia}
\label{Section 3B}
 In the similar fashion of Eq.-(\ref{eqn18a}), we can write the expression for the total moment of inertia $I_{tot}$ as follows,
    \begin{eqnarray}
     && I_{tot}(R)=\int_{V} \rho(r,\theta) (r\sin\theta)^2 r^2\sin\theta dr d\theta d\phi \nonumber\\
     && =\int_{V}\rho^{(0)}(R)(R+\xi)^4 \sin^3\Theta(dR+d\xi) d\Theta d\phi \nonumber\\
     && =\int_{V}\rho^{(0)}(R) R^4 \sin^3\Theta dR d\Theta d\phi \nonumber\\
     && +\int_{V} \rho^{(0)}(R) R^4 \sin^3\Theta \left(\frac{4\xi(R,\Theta)}{R}+\frac{d\xi(R,\Theta)}{dR}\right) dR d\Theta d\phi . \nonumber\\
    \label{eqn31}
    \end{eqnarray}
    The first integral can easily be recognized as the $I^{(0)}(R)$ of Eq.-(\ref{eqn17}). Hence, the total moment of inertia of the rotating star is,
    \begin{eqnarray}
     I_{tot}(a)=I^{(0)}(a)+I^{(2)}(a) \nonumber\\
    \label{eqn32}
    \end{eqnarray}
    where $I^{(2)}(a)$ is the correction in the moment of inertia $\sim \Omega^2$ and can be expressed as,
    \begin{eqnarray}
     I^{(2)}(a)&&=\int\limits_{0}^{a} \int\limits_{0}^{\pi} \int\limits_{0}^{2\pi} \rho^{(0)}(R) R^4 \left[\left(\frac{4\xi_0(R)}{R}+\frac{4\xi_2(R)}{R}P_2(\cos\Theta)\right) \right. \nonumber\\
     &&\;\; +\left. \left(\frac{d\xi_0(R)}{dR}+\frac{d\xi_2(R)}{dR}P_2(\cos\Theta)\right)\right] \sin^3\Theta dR d\Theta d\phi \nonumber\\
     &&=\frac{8\pi}{3}\int_{0}^{a} \rho^{(0)}(R) R^4 \left[\frac{4}{R}\left(\xi_0(r)-\frac{1}{5}\xi_2(R)\right) \right. \nonumber\\
     &&\;\; +\left. \left(\frac{d\xi_0(R)}{dR}-\frac{1}{5}\frac{d\xi_2(R)}{dR}\right)\right] dR  \nonumber\\
     &&=\frac{8\pi}{3}\int_{0}^{a} \left(\frac{1}{5}\xi_2(R)-\xi_0(R)\right) \frac{d\rho^{(0)}(R)}{dR} R^4 dR.
    \label{eqn33}
    \end{eqnarray}
\subsection{Quadrupole Moment}
\label{Section 3C}
    As the potential of the rotating spheroid, $\Phi(R,\Theta)$, is given in Eq.-(\ref{eqn9}), one can write the potential for $R>a$ as follows,
    \begin{equation}
     \Phi(R,\Theta)=\Phi^{(0)}(R)+\Phi^{(2)}_0(R)+\Phi^{(2)}_2(R)P_2(\cos\Theta) .
    \label{eqn34}
    \end{equation}
    The behavior of the terms $\Phi^{(0)}(R)$, $\Phi^{(2)}_0(R)$ for $R>a$ can be found from Eqs.-(\ref{eqn18}) \& (\ref{eqn21}), respectively. On the other hand, for the external behavior of $\Phi^{(2)}_2(R)$ one needs to look into the Eq.-(\ref{eqn25}). Therefore,
    \begin{eqnarray}
     \Phi(R,\Theta)&& =-\frac{GM^{(0)}(a)}{R}-\frac{GM^{(2)}(a)}{R}+\frac{K_1}{R^3}P_2(\cos\Theta) \nonumber\\
     && =-G\frac{M_{tot}}{R}+\frac{K_1}{R^3}P_2(\cos\Theta).
    \label{eqn35}
    \end{eqnarray}
    In the coefficient of the $P_2(\cos\Theta)$, the denominator depends on $R^3$ and therefore, this term represents the contribution of the quadrupole moment in the potential. Hence, the quadrupole moment $Q$ of the rotating star can be defined  as,
    \begin{equation}
     Q=\frac{K_1}{G},
    \label{eqn36}
    \end{equation}
    where $K_1$ has already been fixed from the condition of continuity of $\chi(R)$ and $\Phi^{(2)}_{2}(R)$ at the star's surface in Eq.-(\ref{eqn25}).
\subsection{Rotational Love Number}
\label{Section 3D}
    A constant density spherical surface of radius $R$ becomes oblate shaped spheroid under the action of rotation and therefore, its equatorial radius $r_e(R)$ becomes different from its polar radius $r_p(R)$. This deformation is quantified through ellipticity $\epsilon(R)$ and is defined as \cite{Chandrasekhar1969},
    \begin{eqnarray}
     && \epsilon(R)=\frac{r_e(R)-r_p(R)}{R} \nonumber\\
     \Rightarrow && \epsilon(R)=-\frac{3 \xi_2(R)}{2R}
    \label{eqn37}
    \end{eqnarray}
    This quantity satisfies the following equation \cite{Chandrasekhar1969, Chandrasekhar1963},
    \begin{eqnarray}
     && \frac{M^{0}(R)}{R}\frac{d^2\epsilon(R)}{dR^2}+\frac{2}{R}\frac{dM^{0}(R)}{dR}\frac{d\epsilon(R)}{dR} \nonumber\\
     && +2\frac{dM^{0}(R)}{dR}\frac{\epsilon(R)}{R^2}-6\frac{M^{0}(R)\epsilon(R)}{R^3}=0.
    \label{eqn38}
    \end{eqnarray}
    Introducing the average mass density variable, $\rho_m(R)=3M^{(0)}(R)/4\pi R^3$, for the non-rotating star, it can be shown that, the Eq.-(\ref{eqn38}) takes the form \cite{Tassoul2015},
    \begin{equation}
     R^2\frac{d^2\epsilon(R)}{dR^2}+6\frac{\rho(R)}{\rho_m(R)}\left[R\frac{d\epsilon(R)}{dR}+\epsilon(R)\right]-6\epsilon(R)=0.
    \label{eqn39}
    \end{equation}
    Then Eq.-(\ref{eqn39}) can be shown to take the form,
    \begin{equation}
     R\frac{d\eta_2(R)}{dR}+6D(R)[\eta_2(R)+1]+\eta_2(R)[\eta_2(R)-1]=6,
    \label{eqn40}
    \end{equation}
    where $D(R)=\rho(R)/\rho_m(R)$ and $\eta_2(R)$ is the new variable which is given by,
    \begin{equation}
     \eta_2(R)=\frac{R}{\epsilon(R)}\frac{d\epsilon(R)}{dR}.
    \label{eqn41}
    \end{equation}
    Now, the Eq.-(\ref{eqn40}) is integrated outwards from the center with the  obvious conditions,
    \begin{equation}
     D(R)\rightarrow 1 \;\; \& \;\; \eta_2(R)\rightarrow 0 \;\;\;\; \rm{as} \;\; R\rightarrow 0.
    \label{eqn42}
    \end{equation}
    It is worth mentioning the fact that, the $\eta_2(R)$ turns out to be $\Omega^2$ independent and hence the rotational apsidal constant $k_2$ is defined as,
    \begin{equation}
     k_2=\frac{[3-\eta_2(a)]}{2[2+\eta_2(a)]}.
    \label{eqn43}
    \end{equation}
    Finally, the tidal love number $\lambda$ is defined by the relation,
    \begin{equation}
     \lambda=\frac{2}{3G}a^5 k_2
    \label{eqn44}
    \end{equation}
\subsection{The Angular Velocity $\Omega$}
\label{Section 3E}
    The slow rotation approximation here, is assumed to be valid up to angular velocity $\Omega_K$, the Keplerian angular velocity. It is determined from,
    \begin{equation}
     \Omega_K=\sqrt{\frac{GM_{tot}}{r_e^3}}.
    \label{eqn45}
    \end{equation}
    The initial value for $\Omega_K$ to start the computation is chosen to be the one corresponding to the Keplerian angular velocity of the non-rotating spherical configuration i.e. $\sqrt{GM^{(0)}(a)/a^3}$. After each step of computation $\Omega_K$ is computed from Eq.-(\ref{eqn45}) and using this value of $\Omega_K$ the whole computation is carried out again. Until, a certain level of accuracy is achieved this process goes on and the values of the physical quantities, provided in the final step, are considered for further investigation.
    
\begin{figure}
\vspace{0.0cm}
\eject\centerline{\epsfig{file=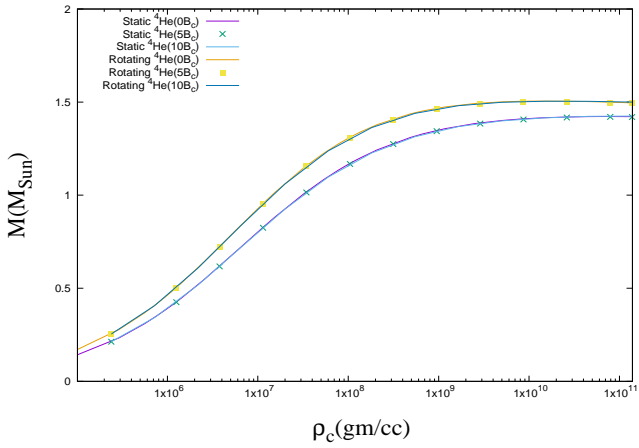,height=6cm,width=9cm}}
\caption{Plots of the mass of the Helium white dwarfs with respect to the central density for different magnetic field strengths.} 
\label{fig1}
\vspace{0.0cm}
\end{figure}

\begin{figure}
\vspace{0.0cm}
\eject\centerline{\epsfig{file=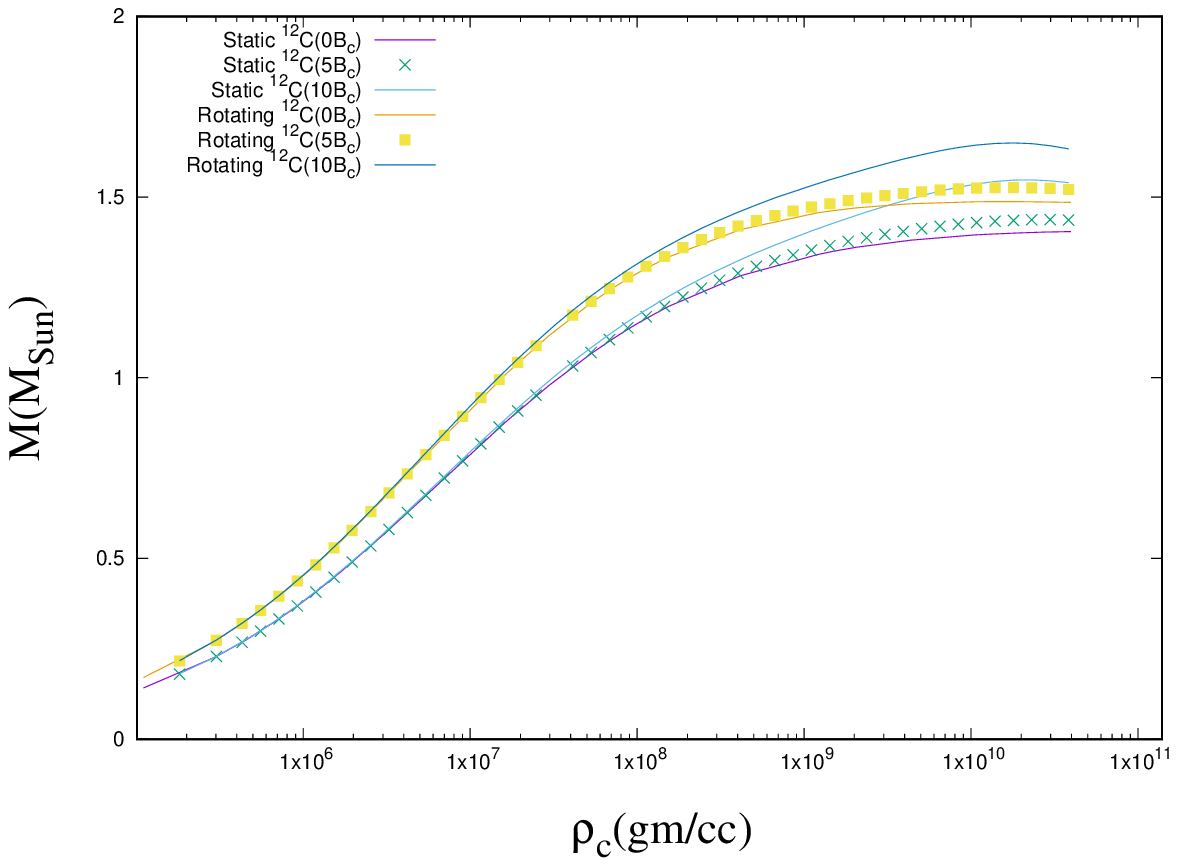,height=6cm,width=9cm}}
\caption{Plots of the mass of the Carbon white dwarfs with respect to the central density for different magnetic field strengths.} 
\label{fig2}
\vspace{0.0cm}
\end{figure}

\begin{figure}
\vspace{0.0cm}
\eject\centerline{\epsfig{file=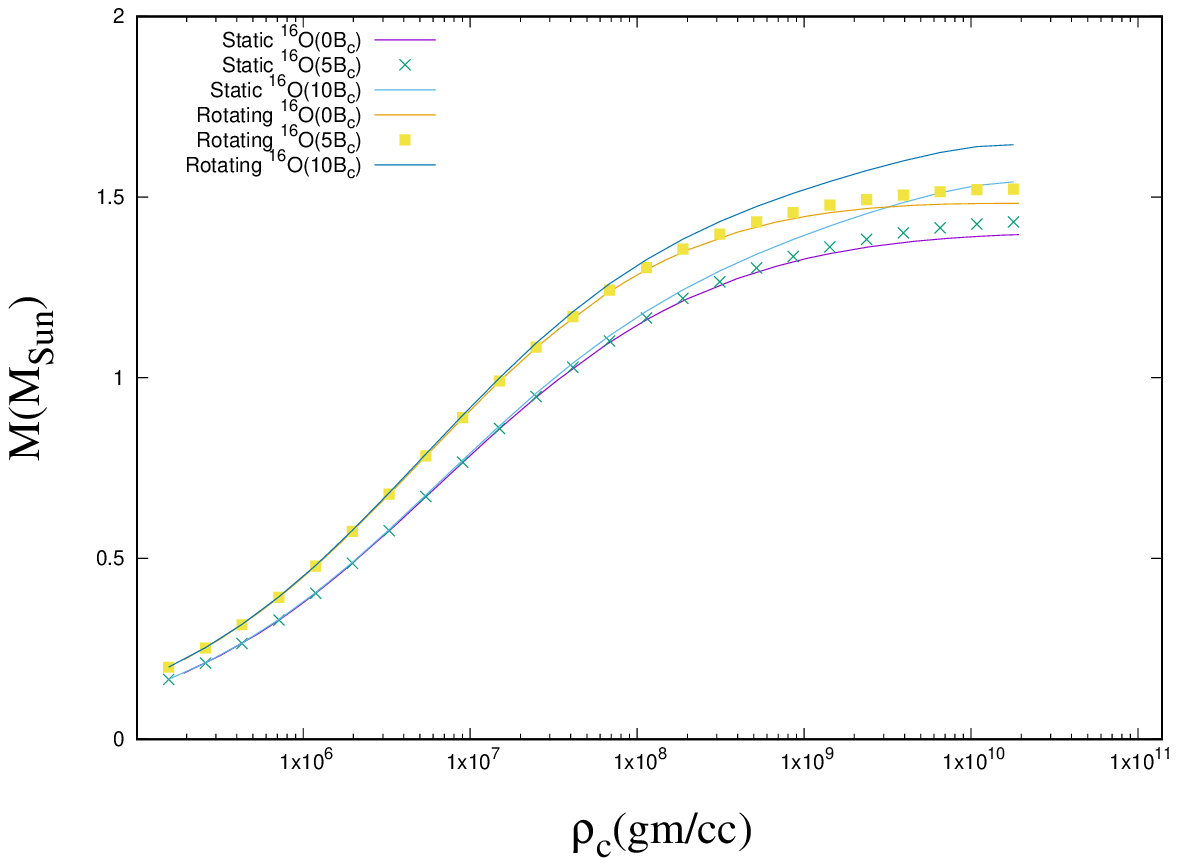,height=6cm,width=9cm}}
\caption{Plots of the mass of the Oxygen white dwarfs with respect to the central density for different magnetic field strengths.} 
\label{fig3}
\vspace{0.0cm}
\end{figure}

\begin{figure}
\vspace{0.0cm}
\eject\centerline{\epsfig{file=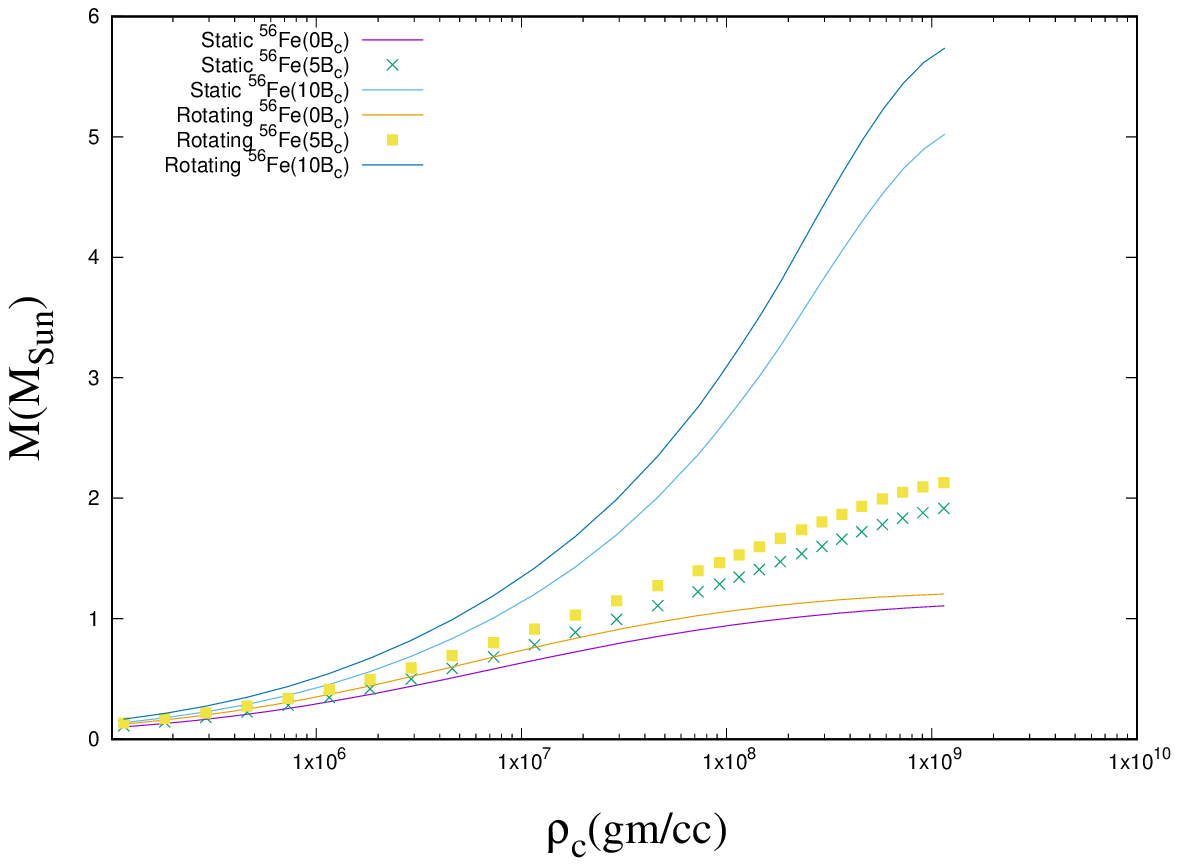,height=6cm,width=9cm}}
\caption{Plots of the mass of the Iron white dwarfs with respect to the central density for different magnetic field strengths.} 
\label{fig4}
\vspace{0.0cm}
\end{figure}

\begin{figure}
\vspace{0.0cm}
\eject\centerline{\epsfig{file=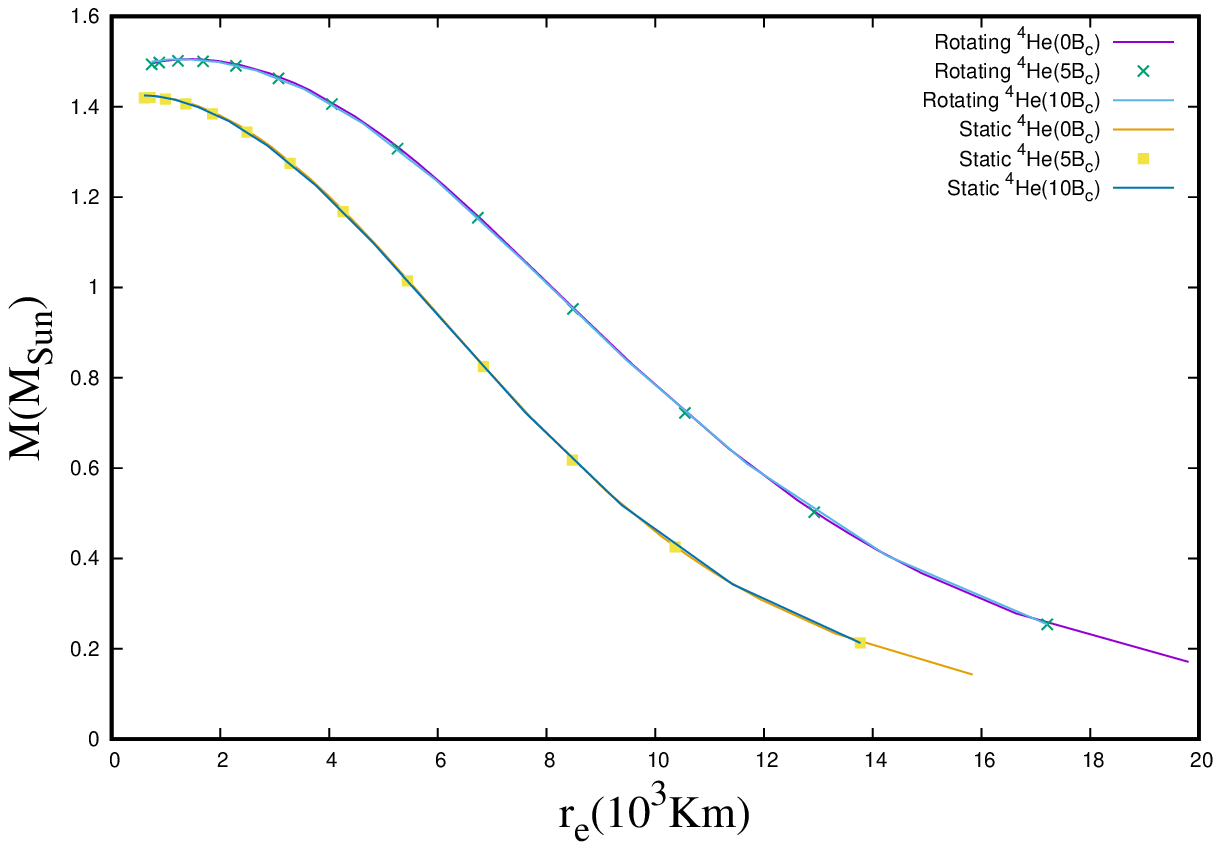,height=6cm,width=9cm}}
\caption{Mass-Radius relationship for Helium white dwarfs for different magnitudes of magnetic field.} 
\label{fig5}
\vspace{0.0cm}
\end{figure}

\begin{figure}
\vspace{0.0cm}
\eject\centerline{\epsfig{file=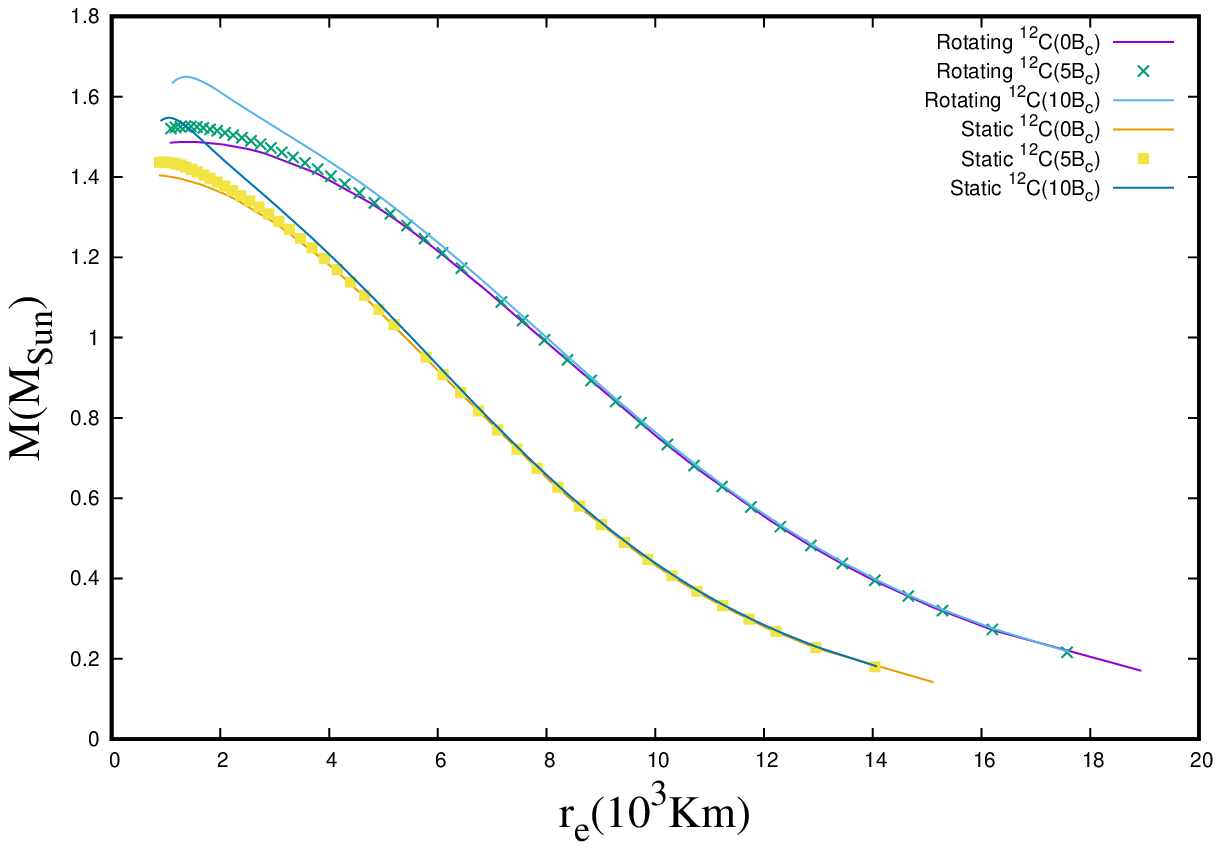,height=6cm,width=9cm}}
\caption{Mass-Radius relationship for Carbon white dwarfs for different magnitudes of magnetic field.} 
\label{fig6}
\vspace{0.0cm}
\end{figure}

\begin{figure}
\vspace{0.0cm}
\eject\centerline{\epsfig{file=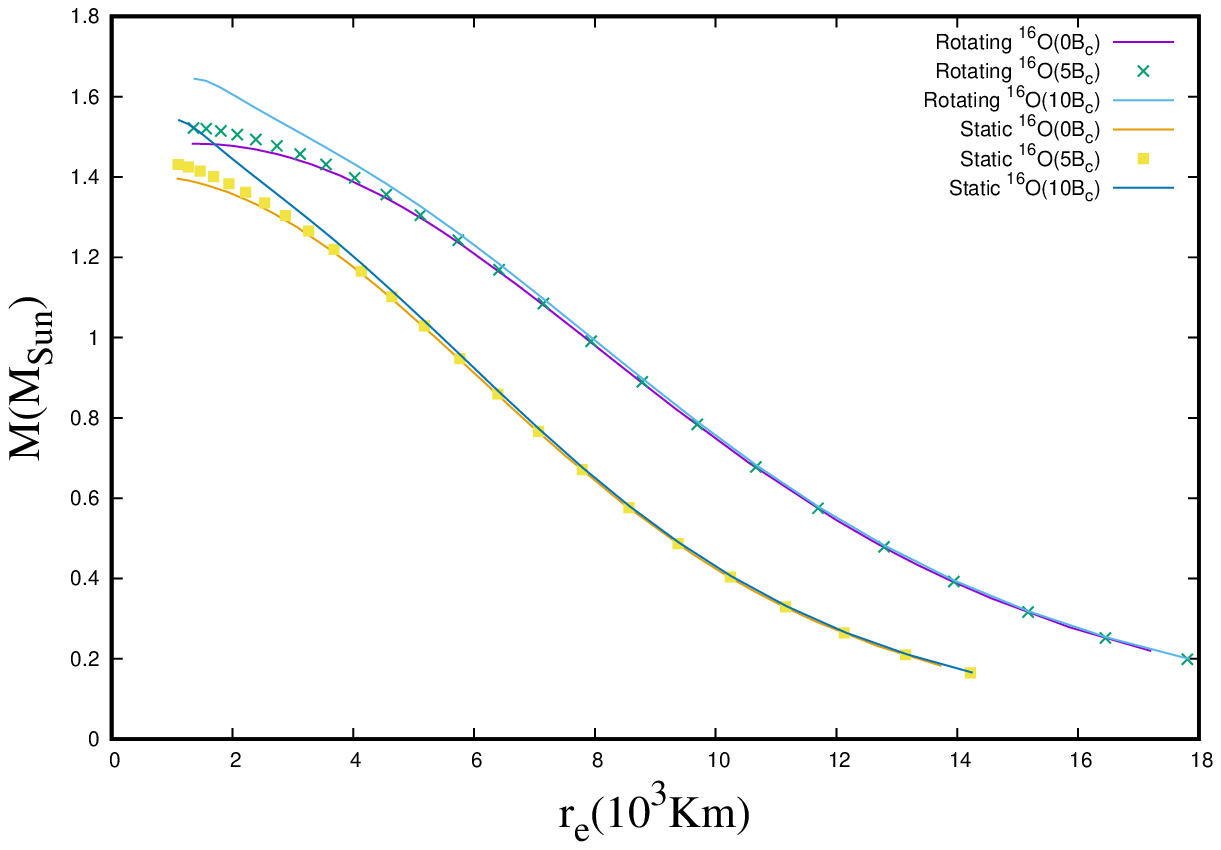,height=6cm,width=9cm}}
\caption{Mass-Radius relationship for Oxygen white dwarfs for different magnitudes of magnetic field.} 
\label{fig7}
\vspace{0.0cm}
\end{figure}

\begin{figure}
\vspace{0.0cm}
\eject\centerline{\epsfig{file=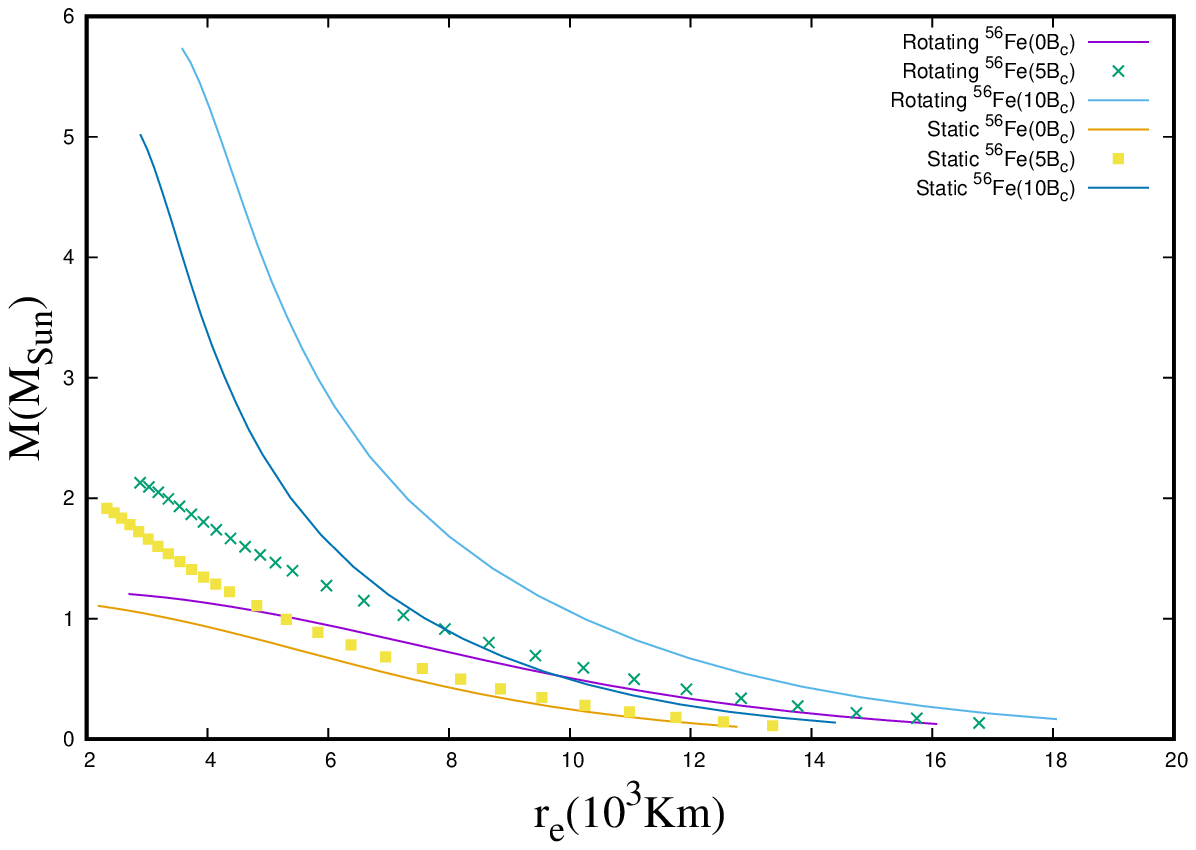,height=6cm,width=9cm}}
\caption{Mass-Radius relationship for Iron white dwarfs for different magnitudes of magnetic field.} 
\label{fig8}
\vspace{0.0cm}
\end{figure}

\begin{figure}
\vspace{0.0cm}
\eject\centerline{\epsfig{file=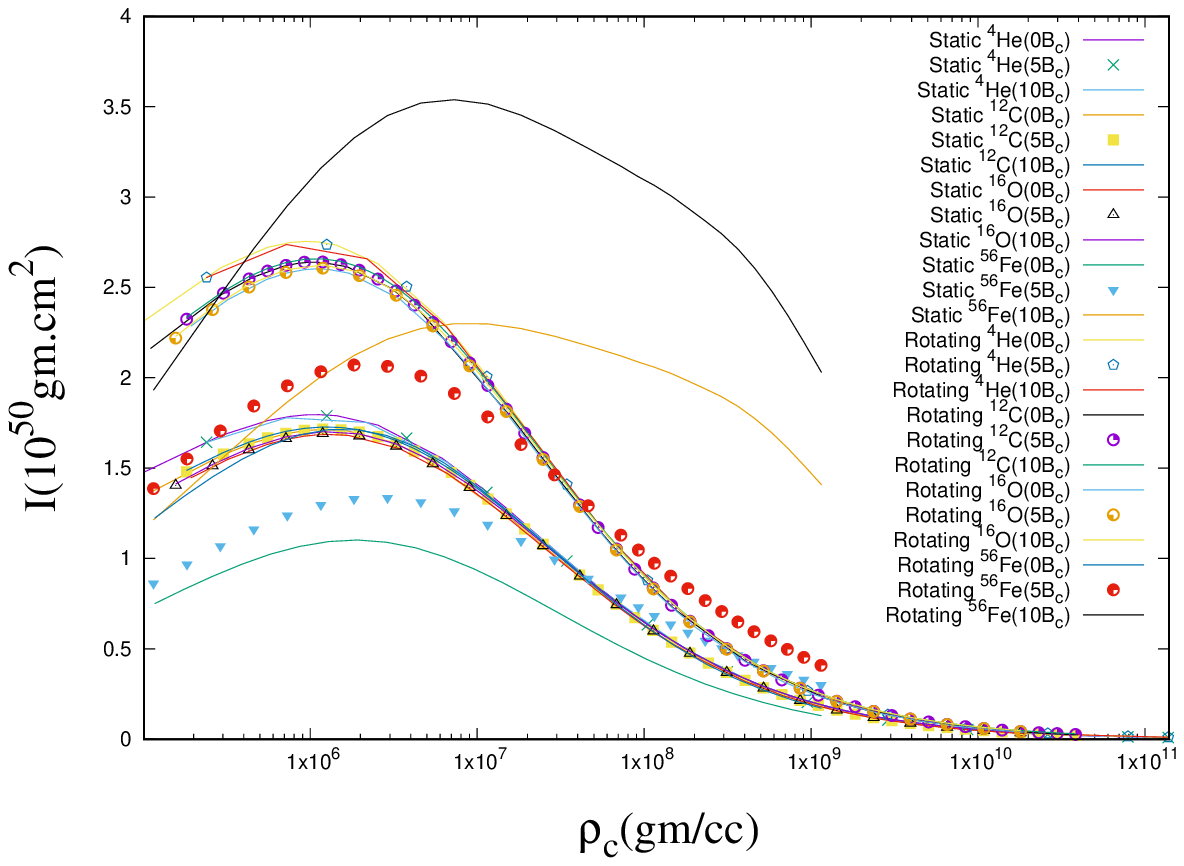,height=6cm,width=9cm}}
\caption{Plots of moment of inertia for Helium, Carbon, Oxygen, Iron white dwarfs with respect to varying central density under different field strengths.} 
\label{fig9}
\vspace{0.0cm}
\end{figure}

\begin{figure}
\vspace{0.0cm}
\eject\centerline{\epsfig{file=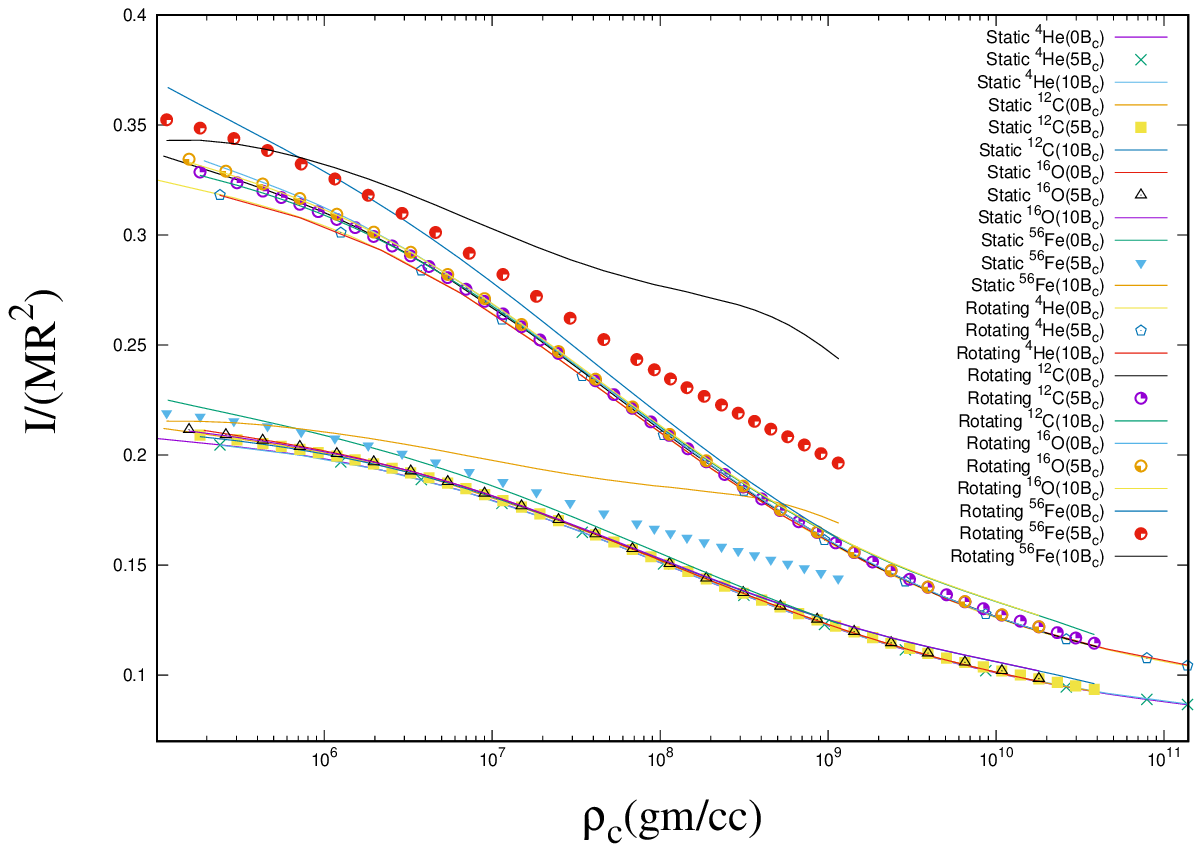,height=6cm,width=9cm}}
\caption{Plots of normalized moment of inertia for Helium, Carbon, Oxygen, Iron white dwarfs with respect to varying central density under different field strengths.} 
\label{fig10}
\vspace{0.0cm}
\end{figure}

\begin{figure}
\vspace{0.0cm}
\eject\centerline{\epsfig{file=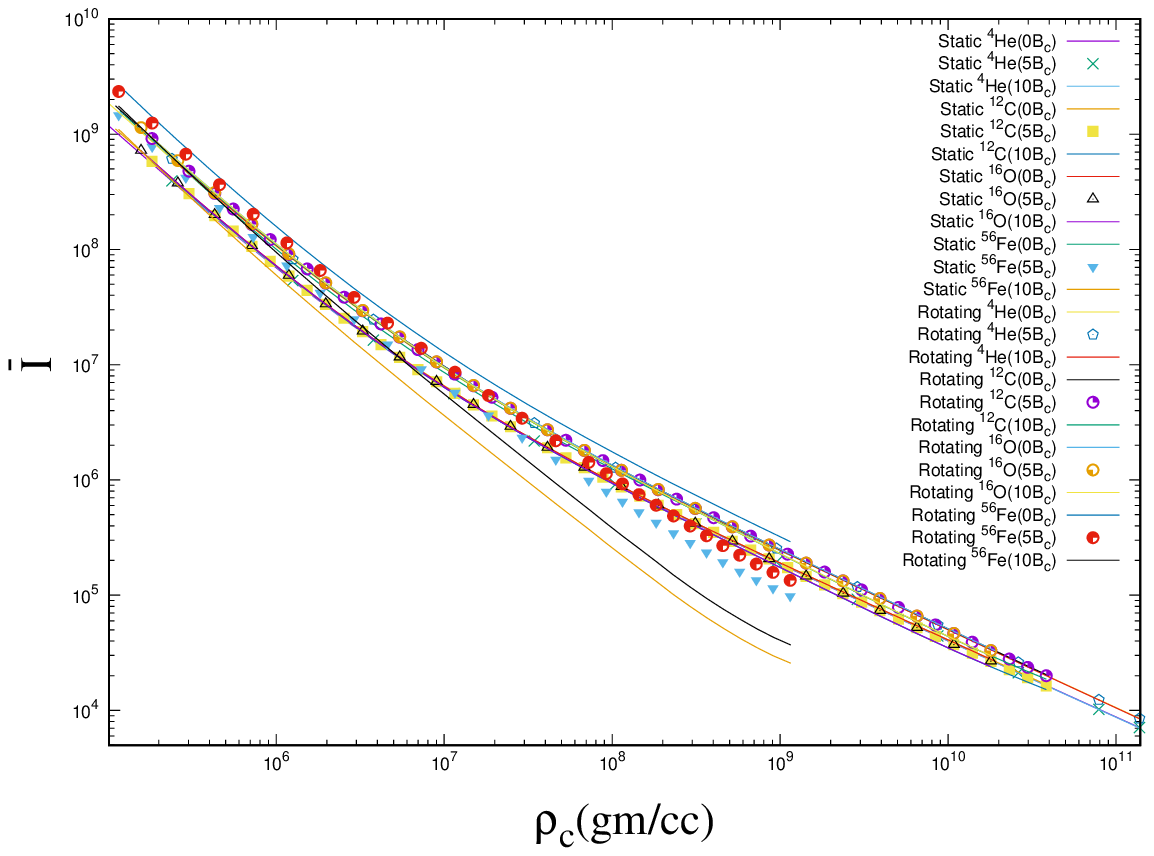,height=6cm,width=9cm}}
\caption{Plots of dimensionless moment of inertia for Helium, Carbon, Oxygen, Iron white dwarfs with respect to varying central density under different field strengths.} 
\label{fig11}
\vspace{0.0cm}
\end{figure}

\begin{figure}
\vspace{0.0cm}
\eject\centerline{\epsfig{file=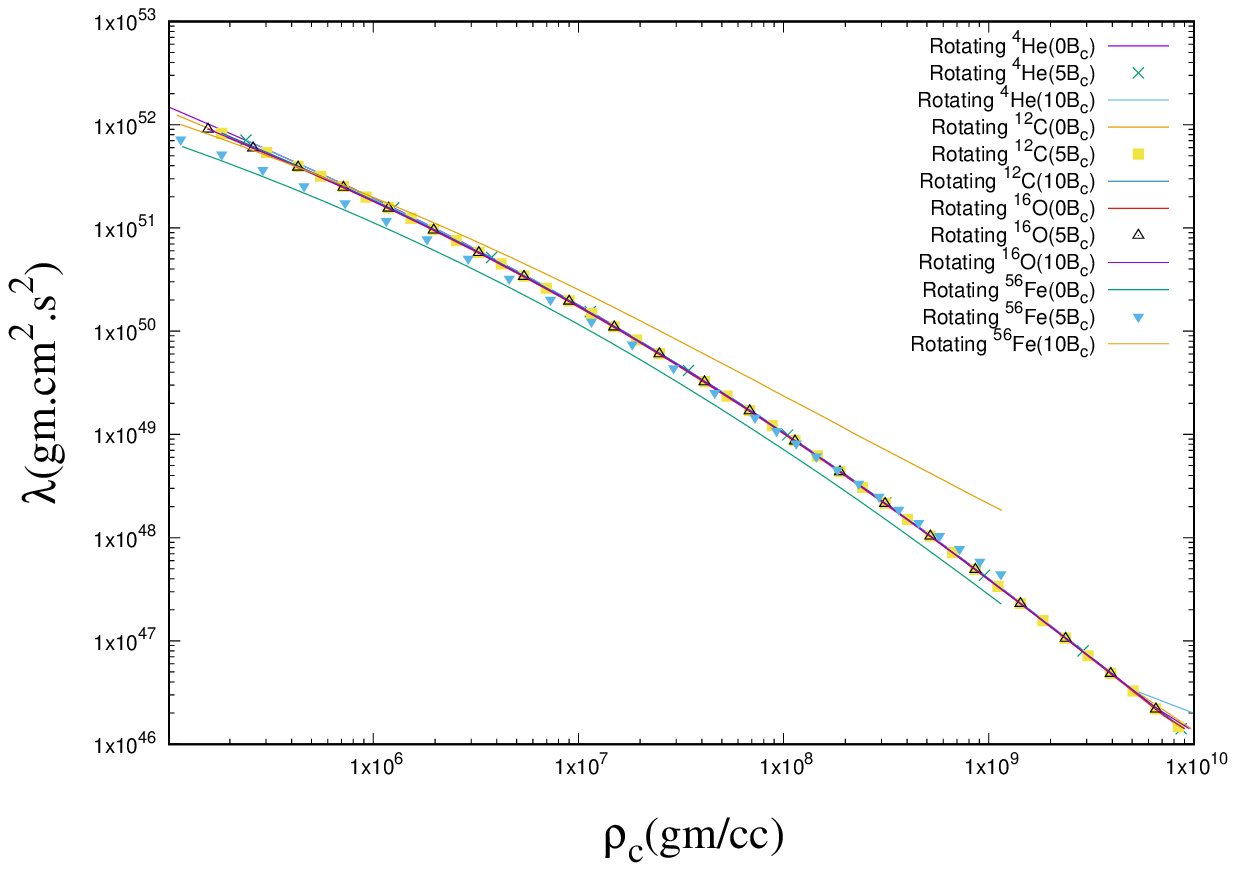,height=6cm,width=9cm}}
\caption{Plots of love number for Helium, Carbon, Oxygen, Iron white dwarfs with respect to varying central density under different field strengths.} 
\label{fig12}
\vspace{0.0cm}
\end{figure}

\begin{figure}
\vspace{0.0cm}
\eject\centerline{\epsfig{file=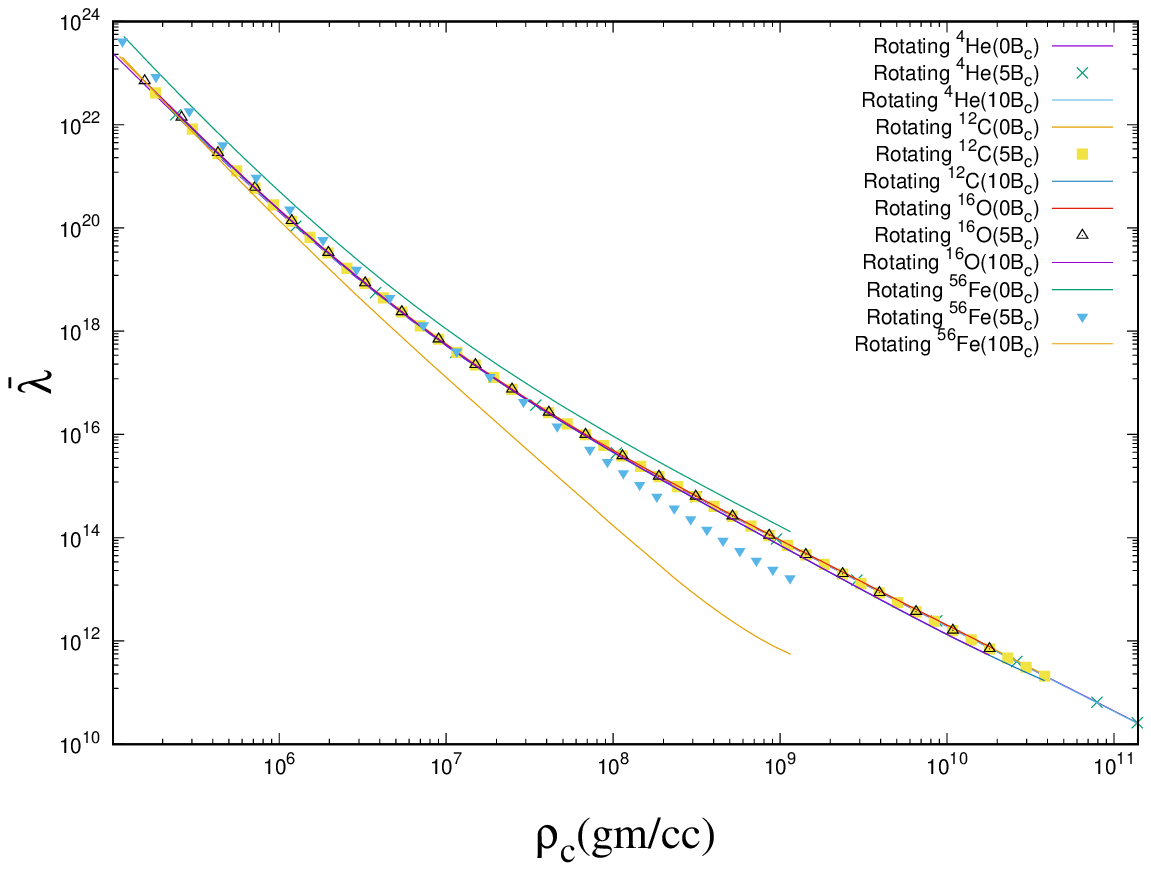,height=6cm,width=9cm}}
\caption{Plots of dimensionless love number for Helium, Carbon, Oxygen, Iron white dwarfs with respect to varying central density under different field strengths.} 
\label{fig13}
\vspace{0.0cm}
\end{figure}

\begin{figure}
\vspace{0.0cm}
\eject\centerline{\epsfig{file=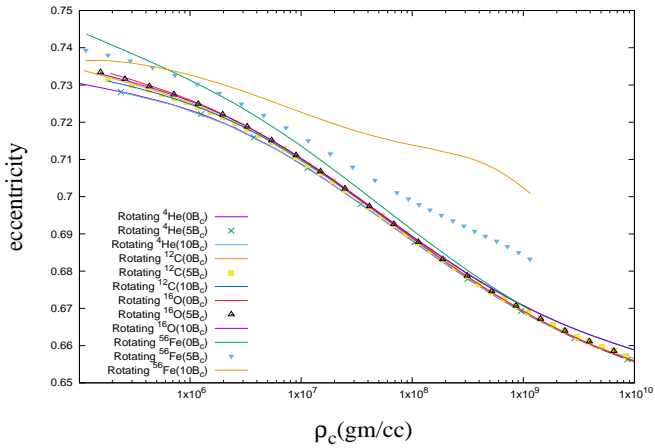,height=6cm,width=9cm}}
\caption{Plots of eccentricity for Helium, Carbon, Oxygen, Iron white dwarfs with respect to varying central density under different field strengths.} 
\label{fig14}
\vspace{0.0cm}
\end{figure}

\begin{figure}
\vspace{0.0cm}
\eject\centerline{\epsfig{file=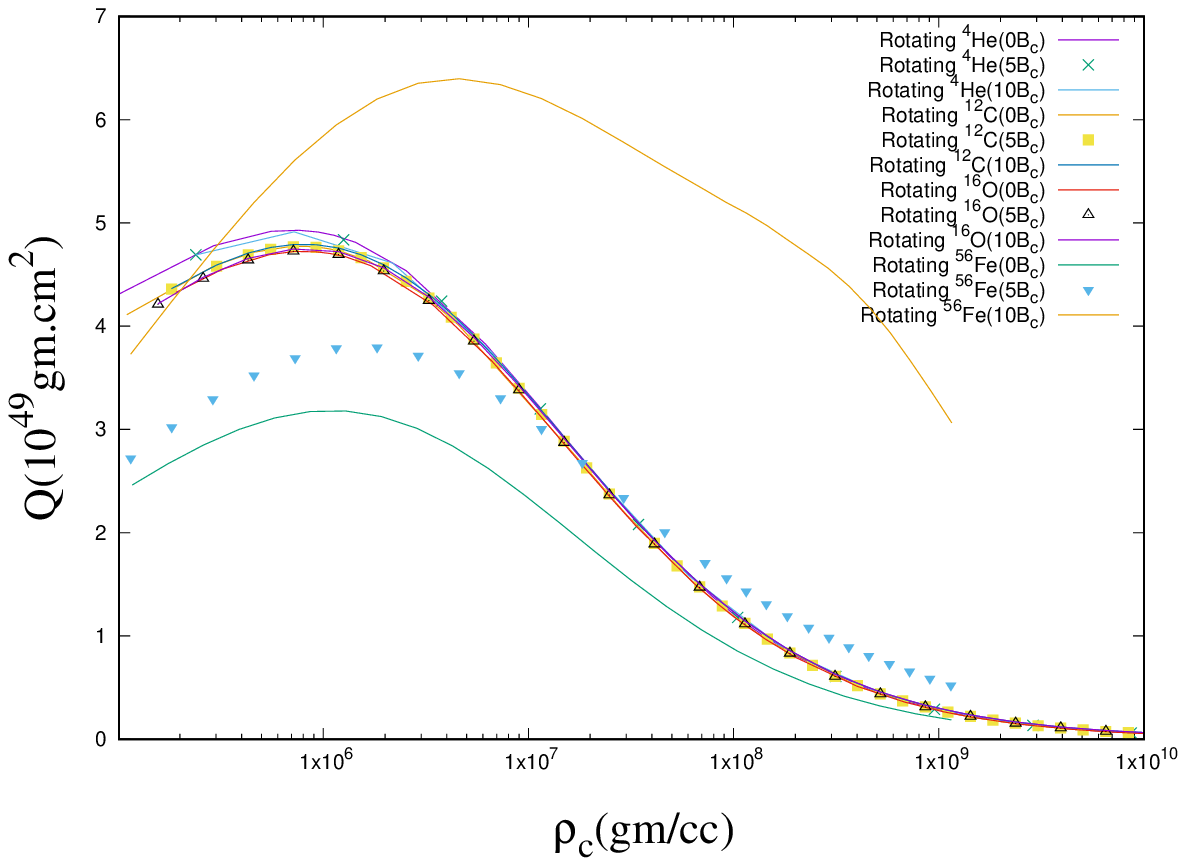,height=6cm,width=9cm}}
\caption{Plots of spin quadrupole moment for Helium, Carbon, Oxygen, Iron white dwarfs with respect to varying central density under different field strengths.} 
\label{fig15}
\vspace{0.0cm}
\end{figure}

\begin{figure}
\vspace{0.0cm}
\eject\centerline{\epsfig{file=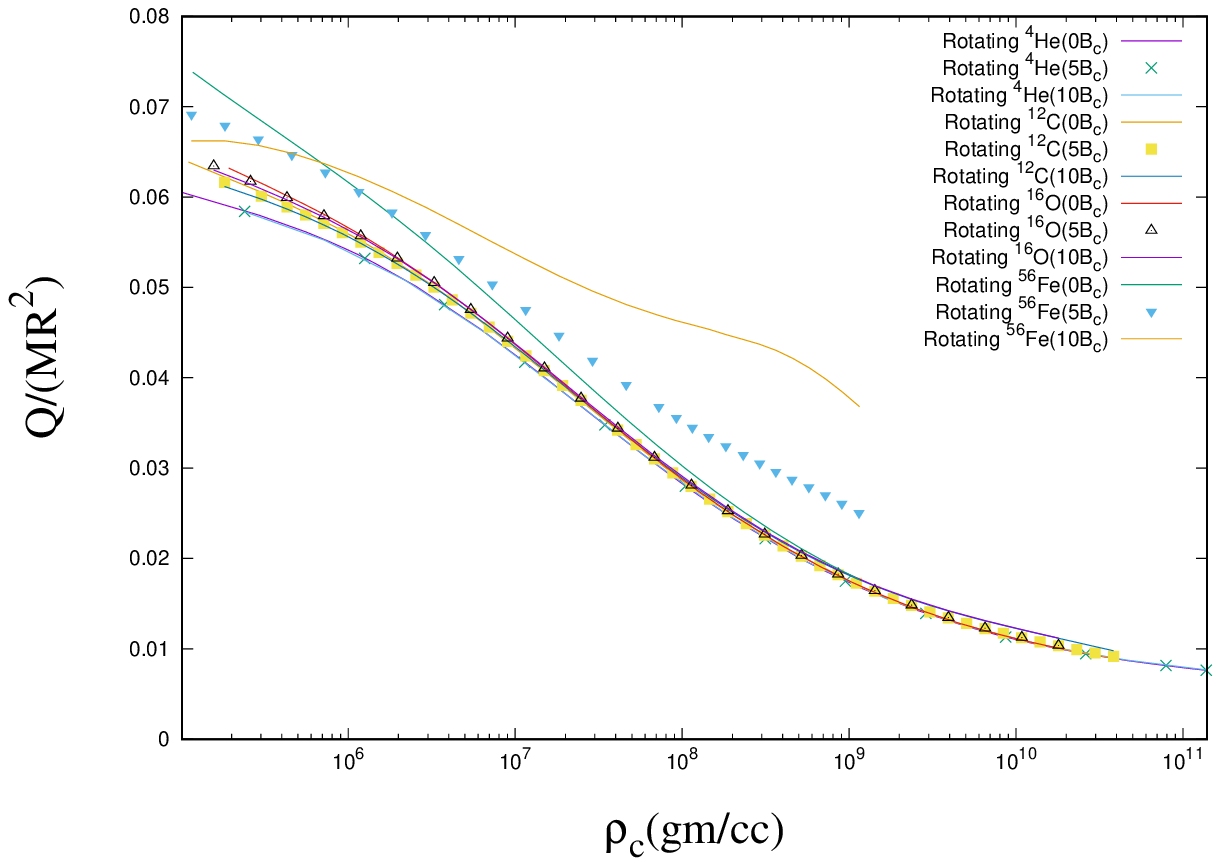,height=6cm,width=9cm}}
\caption{Plots of normalized spin quadrupole moment for Helium, Carbon, Oxygen, Iron white dwarfs with respect to varying central density under different field strengths.} 
\label{fig16}
\vspace{0.0cm}
\end{figure}

\begin{figure}
\vspace{0.0cm}
\eject\centerline{\epsfig{file=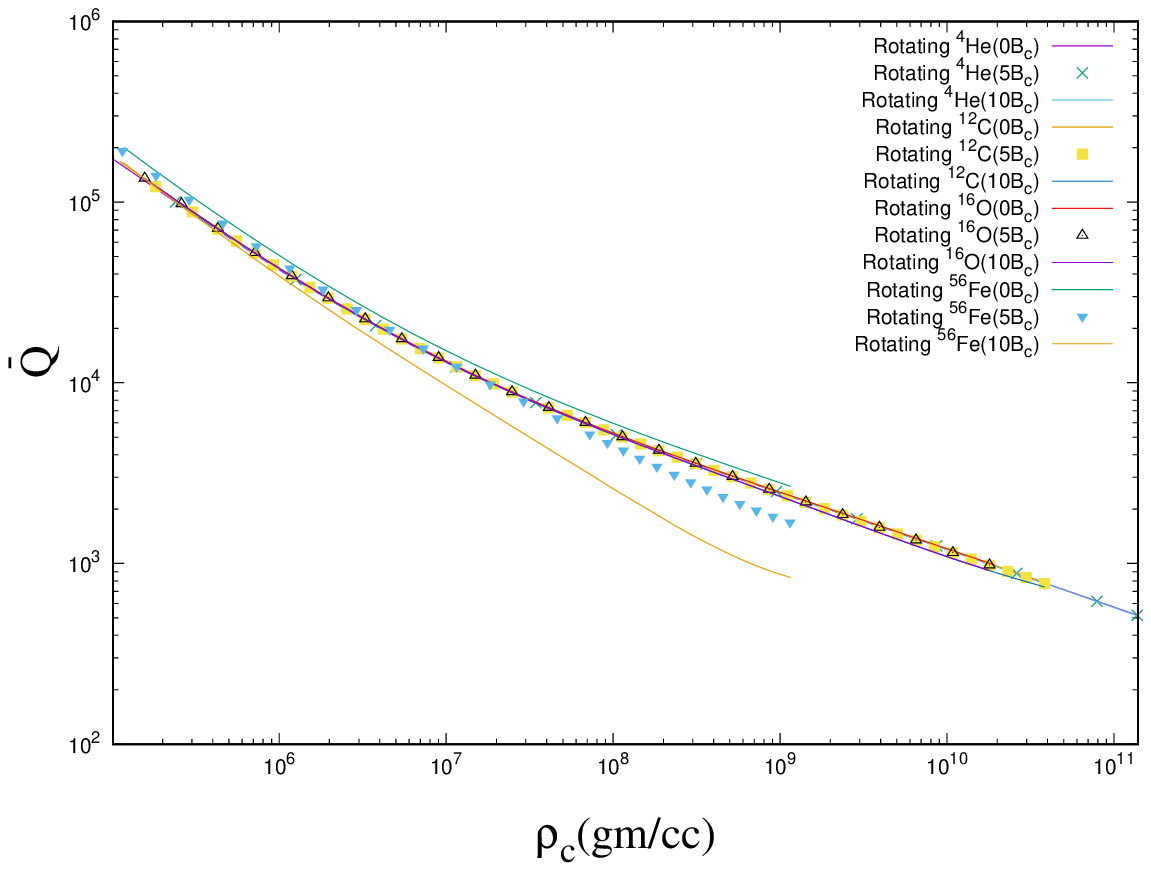,height=6cm,width=9cm}}
\caption{Plots of dimensionless spin quadrupole moment for Helium, Carbon, Oxygen, Iron white dwarfs with respect to varying central density under different field strengths.} 
\label{fig17}
\vspace{0.0cm}
\end{figure}

\begin{figure}
\vspace{0.0cm}
\eject\centerline{\epsfig{file=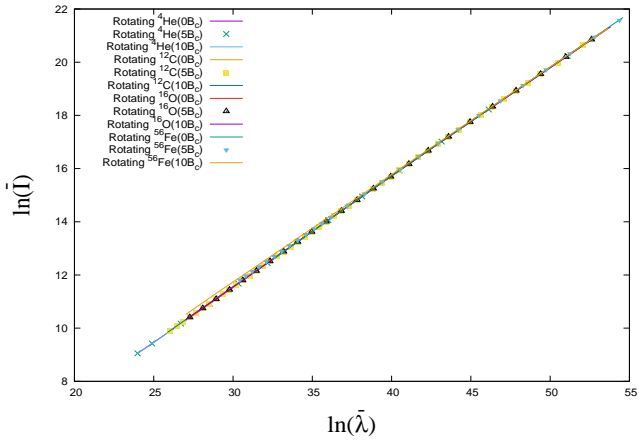,height=6cm,width=9cm}}
\caption{Plots of logarithmic dimensionless moment of inertia with respect to logarithmic dimensionless love number for different white dwarfs and different field strengths.} 
\label{fig18}
\vspace{-0.4cm}
\end{figure}

\begin{figure}
\vspace{0.0cm}
\eject\centerline{\epsfig{file=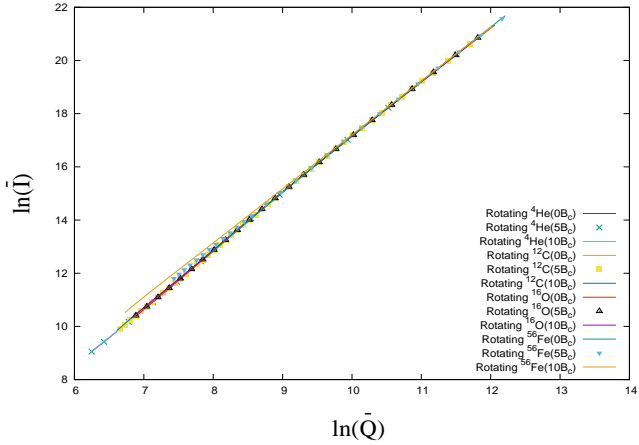,height=6cm,width=9cm}}
\caption{Plots of logarithmic dimensionless moment of inertia with respect to logarithmic dimensionless quadrupole moment for different white dwarfs and different field strengths.} 
\label{fig19}
\vspace{0.0cm}
\end{figure}

\begin{figure}
\vspace{0.0cm}
\eject\centerline{\epsfig{file=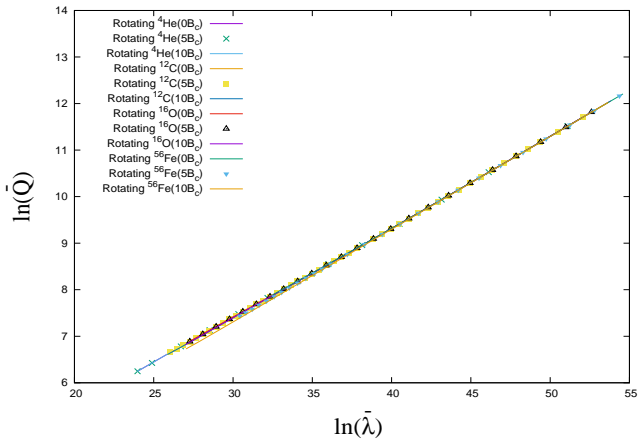,height=6cm,width=9cm}}
\caption{Plots of logarithmic dimensionless quadrupole moment with respect to logarithmic dimensionless love number for different white dwarfs and different field strengths.} 
\label{fig20}
\vspace{0.0cm}
\end{figure}

\begin{figure}
\vspace{0.0cm}
\eject\centerline{\epsfig{file=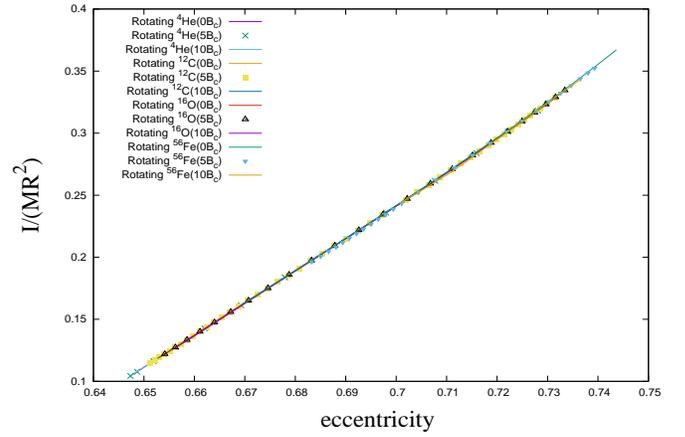,height=6cm,width=9cm}}
\caption{Plots of normalized moment of inertia with respect to eccentricity for different white dwarfs and different field strengths.} 
\label{fig21}
\vspace{0.0cm}
\end{figure}

\begin{figure}
\vspace{0.0cm}
\eject\centerline{\epsfig{file=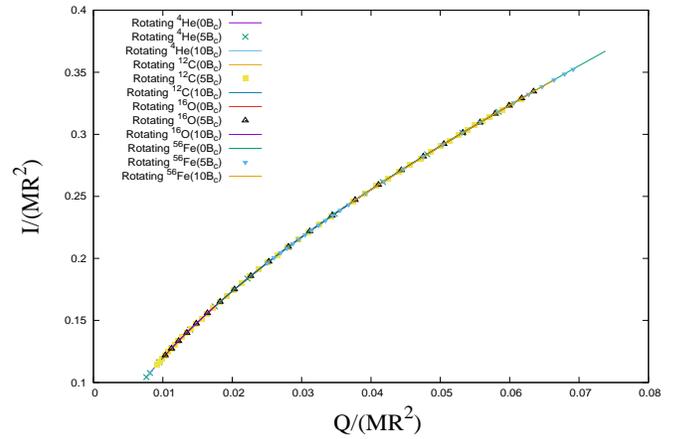,height=6cm,width=9cm}}
\caption{Plots of normalized moment of inertia with respect to normalized quadrupole moment for different white dwarfs and different field strengths.} 
\label{fig22}
\vspace{0.0cm}
\end{figure}

\begin{figure}
\vspace{0.0cm}
\eject\centerline{\epsfig{file=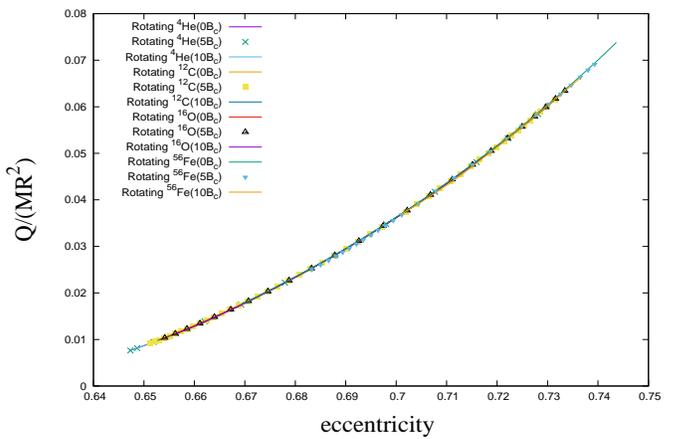,height=6cm,width=9cm}}
\caption{Plots of normalized quadrupole moment with respect to eccentricity for different white dwarfs and different field strengths.} 
\label{fig23}
\vspace{-0.4cm}
\end{figure}

\noindent
\section{The equation of state}
\label{Section 4}
    The equation of state that we adopted here is the one developed in Ref.-\cite{Roy2019} i.e. the extension of the Feynman-Metropolis-Teller(FMT) treatment to treat magnetized super-Chandrasekhar white dwarfs. For this each of the atomic configuration with atomic number $Z$ and mass number $A$ has been considered as Wigner-Seitz cell and the electrons within the cell occupy Landau quantized states when subjected to magnetic field $B$ and the maximum number of particles per Landau level per unit area is $\frac{eB(2s+1)}{hc}$. When the magnetic field $B$ is in z-direction the Fermi energy $E_F$ of the electron at $\nu $th Landau level is given by, 
    
\vspace{0.cm}
\begin{equation}
\vspace{0.cm}
E_F=\left[p_F^2(\nu)c^2+m_e^2c^4\left(1+2\nu B_D\right)\right]^\frac{1}{2}-m_ec^2-eV(r)
\label{seqn22}
\vspace{0.0cm}
\end{equation}
\noindent
where $e$ is the electronic charge, $m_e$ is the electronic rest mass, $p_{F}$ is the Fermi momentum of the electron, $c$ is the speed of light and $V(r)$ is the Coulomb potential at radius $r$. The term $B_D=B/B_c$, with $B_c$ being the magnetic field corresponding to which the Landau quantization energy becomes equal to that of the electron rest mass energy.

    The number density of electrons under the influence of Coulomb screening is given by,

\vspace{0.cm}
\begin{eqnarray}
\vspace{0.cm}
n_e(r)=&&\frac{2B_D}{(2\pi )^2\lambda_e^3}\sum\limits_{\nu =0}^{\nu_m} g_{\nu}x_F(\nu)
\label{seqn23}
\vspace{0.cm}
\end{eqnarray}
\noindent
where $g_{\nu}$ is the degeneracy of the $\nu$th Landau level and $x_F(\nu)=\frac{p_F(\nu )c}{m_ec^2}$. $\nu_m$ is the highest occupied landau level determined from positive semi-definiteness of $p_F(\nu)$. The overall Coulomb potential $V(r)$ can be obtained by solving the Poisson equation

\vspace{0.cm}
\begin{eqnarray}
\vspace{0.cm}
\nabla^2V(r)=-4\pi e[n_p(r)-n_e(r)] \nonumber\\
\Rightarrow \nabla^2\widehat{V}(r)=-4\pi e^2[n_p(r)-n_e(r)] 
\label{seqn24}
\vspace{0.cm}
\end{eqnarray}
\noindent
where $n_p(r)=3Z/4\pi R_c^3$ is the constant proton density within the nuclear radius $R_c$ and $\hat{V}(r)=eV(r)+E_F$.

    Introducing the dimensionless quantities $x=\frac{r}{\lambda_\pi}$ and $y(x)= r\frac{\widehat{V}(r)}{\hbar c}=x\frac{\widehat{V}(x)}{m_\pi c^2}$, the Eq.(\ref{seqn24}) can be rewritten as, 
    
\begin{eqnarray}
&& \frac{1}{x}\frac{d^2y(x)}{dx^2}= -\frac{3\alpha \theta(x_c-x)}{\Delta^3} +\frac{2 e^2B_D}{\pi \hbar c}\left(\frac{m_\pi}{m_e}\right)\left(\frac{\lambda_\pi}{\lambda_e}\right)^3 \times \nonumber\\ 
&& \sum\limits_{\nu =0}^{\nu_m} g_{\nu}\left[\left(1-\frac{\nu}{\nu_m}\right)\left\{\left(\frac{y}{x}\right)^2+2\left(\frac{m_e}{m_\pi}\right)\left(\frac{y}{x}\right)\right\}\right]^\frac{1}{2},
\label{seqn25}
\vspace{0.cm}
\end{eqnarray}
\noindent
where $\hbar$ is the Planck's constant, $\alpha=e^2/\hbar c$ is the fine structure constant, $\Delta=R_c/\lambda_{\pi}Z^{\frac{1}{3}}$, $\theta(x_c-x)$ is the Heavyside step function with $x_c=R_c/\lambda_{\pi}$, $m_{\pi}$ is the pion rest mass, $\lambda_{\pi}=\hbar c/m_{\pi}c^2$ is the Compton wavelength for pion and the Compton wavelength for electron is $\lambda_{e}=\hbar c/m_{e}c^2$. The Eq.(\ref{seqn25}) is integrated numerically to find out the electronic distribution within the cell and other subsequent quantities where the constraints and boundary conditions arise out of the physical requirements of the system itself. More details on the numerical integration and electron distribution within the Wigner-Seitz cell can be found on \cite{Roy2019,Ahmad2020} and references therein.

    The kinetic energy density including the electronic rest mass within the Wigner-Seitz cell then turns out to be,
\begin{equation}
\varepsilon_k (x)=\frac{B_Dm_ec^2}{2\pi^2\lambda_e^3} \sum\limits_{\nu =0}^{\nu_m} g_\nu\left(1+2\nu B_D\right) 
\psi \left(\frac{x_F(\nu )}{(1+2\nu B_D)^{1/2}}\right)  \nonumber\\
\label{seqn26}
\vspace{0.0cm}
\end{equation}
\noindent
where it should be noted that,
\begin{eqnarray}
 \psi (z)=&& \int_0^z(1+y^2)^{1/2}dy \nonumber\\
 =&& \frac{1}{2}[z\sqrt{1+z^2}+\ln(z+\sqrt{1+z^2})].
\label{seqn26a}
\end{eqnarray}
The total kinetic energy $E_k$ of the cell excluding the electronic rest mass can be calculated as,
    
\vspace{0.cm}
\begin{equation}
\vspace{0.cm}
E_k=4\pi\lambda_\pi^3\int_{x_c}^{x_{WS}} x^2[\varepsilon_k (x)-n_e(x)m_ec^2]dx
\label{seqn27}
\vspace{0.0cm}
\end{equation}
\noindent
where $R_{WS}$ is the radius of the Wigner-Seitz cell and $x_{WS}=R_{WS}/\lambda_{\pi}$.

    The total potential energy $E_c$ of the cell can be evaluated using,
    
\vspace{0.cm}
\begin{eqnarray}
\vspace{0.cm}
&& E_c=4\pi\lambda_\pi^3\int_0^{x_{WS}} x^2[n_p(x)-n_e(x)]eV(x)dx \nonumber \\
\Rightarrow && E_c=-4\pi\lambda_\pi^3\int_{x_c}^{x_{WS}} x^2n_e(x)eV(x)dx \\
\Rightarrow && E_c=-4\pi\lambda_\pi^3 m_\pi c^2 \int_{x_c}^{x_{WS}} xn_e(x) y(x)dx +E_F Z. \nonumber
\label{seqn28}
\vspace{0.0cm}
\end{eqnarray}
\noindent

    The energy density $\varepsilon$ can now be given by

\vspace{0.0cm}
\begin{equation}
\vspace{0.cm}
\varepsilon=\frac{E_k+E_c+M(A,Z)c^2}{\frac{4\pi}{3}R_{WS}^3}+\frac{B^2}{8\pi} 
\label{seqn29}
\vspace{0.cm}
\end{equation}
\noindent    
where $M(A,Z)$ is atomic mass of the uncompressed atom and the last term accounts for the magnetic contribution to the energy density. The pressure $P$ is simply given by,

\vspace{0.cm}
\begin{eqnarray}
\vspace{0.cm}
P=&& \left\lbrace \frac{B_Dm_ec^2}{2\pi^2\lambda_e^3}\sum\limits_{\nu =0}^{\nu_m} g_\nu\left(1+2\nu B_D\right)\eta \left(\frac{x_F(\nu )}{(1+2\nu B_D)^{1/2}}\right) \right. \nonumber\\
&&\left. ~ +\frac{B^2}{24\pi} \right\rbrace_{x_{WS}}
\label{seqn30}
\vspace{0.cm}
\end{eqnarray}
\noindent
where the function $\eta$ is defined as,
\begin{eqnarray}
 \eta (z)=&& z\sqrt{1+z^2}-\psi (z) \nonumber\\
 =&& \frac{1}{2}[z\sqrt{1+z^2}-\ln(z+\sqrt{1+z^2})].
\label{seqn31}
\end{eqnarray}

\subsection{Onset of $\beta$-instability}
\label{Section 4B}
   High central density may lead to the onset of electron capturing by the nucleus  and hence, the inverse $\beta$-decay sets in. In this work the highest value for the central density $\rho_c$ has been chosen to be the critical limit where the inverse $\beta$-decay just starts. The limits for $^4$He, $^{12}$C, $^{16}$O and $^{56}$Fe have been tabulated in Table-\ref{table2}, where $\epsilon_\beta(Z)$ is the experimental inverse $\beta$-decay energy \cite{Roy2019,Ro11,Au03,Au77,Sh83}, $\rho^{\beta,unif}_{crit}$ is the critical density calculated assuming uniform electron distribution inside Wigner-Seitz cell and $\rho^{\beta,relFMT}_{crit}$ is the one calculated under magnetic field assuming non-uniform electron distribution.
   
\begin{table}[htbp]
\centering
\caption{The critical limits for the onset of $\beta$-instability.}
\begin{tabular}{||c|c|c|c||}
\hline 
\hline
Decay channel&$\epsilon_\beta(Z)$&$\rho^{\beta,relFMT}_{crit}$(5B$_c$)&$\rho^{\beta,unif}_{crit}$ \\ \hline
&MeV& g cm$^{-3}$ &g cm$^{-3}$ \\ \hline
\hline
$^4$He$\rightarrow$ $^3$H+n$\rightarrow$ 4n &20.596&1.368$\times$10$^{11}$&1.37$\times$10$^{11}$ \\
$^{12}$C$\rightarrow$ $^{12}$B$\rightarrow$ $^{12}$Be&13.370&3.812$\times$10$^{10}$&3.88$\times$10$^{10}$ \\
$^{16}$O$\rightarrow$ $^{16}$N$\rightarrow$ $^{16}$C &10.419&1.785$\times$10$^{10}$&1.89$\times$10$^{10}$ \\
$^{56}$Fe$\rightarrow$ $^{56}$Mn$\rightarrow$ $^{56}$Cr&3.695&1.139$\times$10$^{9}$&1.14$\times$10$^{9}$ \\ \hline
\hline
\end{tabular}
\label{table2} 
\end{table}
   
\subsection{Magnetic Field}
\label{Section 4A}
    On the basis of the discussion provided in Ref.-\cite{Roy2019} the actual calculations have been performed with varying magnetic field including the effects of energy density and pressure arising due to magnetic field. The density dependent magnetic field \cite{Ba97} inside white dwarf is taken to be of the form,   

\begin{equation}
 B_D = B_s + B_0[1-\exp\{-\beta(\rho/\rho_0)^\gamma\}]
\label{seqn32}
\end{equation} 
\noindent
where $\rho_0$ is taken as $\rho^{\beta,relFMT}_{crit}$/10 and $\beta$, $\gamma$ are constants. We choose constants $\beta=0.8$ and $\gamma=0.9$ to ensure the flatness of the field near surface and center. For the present calculation the magnetic field $B_D$ at center of the white dwarf has been kept up to 10$B_c$ which is $4.414 \times 10^{14}$ gauss \cite{Ch13,Ch15}. The parameter value $B_0$ is fixed by setting a particular value for the magnetic field $B_D$ at the center with the maximum central density i.e. $\rho^{\beta,relFMT}_{crit}$. The surface magnetic field $B_s$ $\sim10^{9}$ gauss estimated by observations \cite{Ke13,Ke15,Fe15}.
                     
\noindent
\section{Results, Discussion and Conclusion}
\label{Section 6}

   In equilibrium configuration the gravitational pull is balanced by the pressure and centrifugal force due to rotation. The Figs.-\ref{fig1}-\ref{fig4} are showing mass variations of the equilibrium configurations with the central density for Helium, Carbon, Oxygen, Iron white dwarfs, respectively. Here the rotational frequency of the star has been taken to be the one determined from Eq.-(\ref{eqn45}), i.e. the Keplerian angular velocity. Magnetic field profile under which the calculations of equilibrium configurations have been carried out is given by Eq.-(\ref{seqn32}). The central magnetic field for each of the configuration is the one found by setting the $\rho$ in Eq.-(\ref{seqn32}) equal to the central density for the configuration and other parameters of the equation are determined following the description in Section-\ref{Section 4A}. It can be seen from these figures that, the change in the critical mass due to the presence of magnetic field is least for $^4$He white dwarfs and highest for the $^{56}$Fe white dwarfs. The rotation has its common effect in increasing the critical mass. The decrement in the central density shows continuous decrease of the mass of the equilibrium configurations irrespective of any applied magnetic field strength. Unlike the general relativistic case $^4$He and $^6$C white dwarfs show no peak in the mass-central density plots. Therefore, having no secular instabilities for rest of the calculations the $\rho^{\beta,relFMT}_{crit}$ has been taken as the highest limit for the central density for all the elements.
   
   Figs.-\ref{fig5}-\ref{fig8} are depicting the corresponding mass-radius relationships for $^{4}$He, $^{12}$C, $^{16}$O and $^{56}$Fe white dwarfs, respectively. Applied magnetic field and other parameters are the same as those used for Figs.-\ref{fig1}-\ref{fig4}. These figures are again showing the minimum effect of the magnetic field on the $^4$He white dwarf mass-radius relationship and the maximum effect on the $^{56}$Fe white dwarf mass-radius relationship. According to the Eq.-(\ref{seqn32}) and the description provided in the Section-\ref{Section 4A}, the central magnetic field is lower for lower central density and consequently it has been found that the mass-radius relationship is more or less unaltered in the lower central density regime irrespective of the field strength. Only the $^{56}$Fe white dwarf in Fig.-\ref{fig8} retains some difference in the mass-radius relationship in the lower density region. This in turn proving the validity of the statement proposed in section-\ref{Section 1} that, the magnetic field pressure is small compared to the matter pressure and hence the pressure splitting due to magnetic field as well.
   
    Figs.-\ref{fig9}-\ref{fig11} shows the moment of inertia counterpart for all the elements, where along y axes moment of inertia in unit of 10$^{50}$ gm.cm$^2$, normalized moment of inertia, dimensionless moment of inertia have been plotted, for both the cases of rotating and non-rotating stars. In Fig.-\ref{fig9} all the curves assume more or less the same feature of initially increasing with increasing density and then falling, whereas, in Figs.-\ref{fig10} \& \ref{fig11} all the curves fall smoothly when central density is increased. The normalization of moment of inertia has been done by dividing the moment of inertia by MR$^2$, where M and R are the mass and equatorial radius of the non-rotating configuration, respectively. In Fig.-\ref{fig11} the dimensionless moment of inertia is given by,
    
    \begin{equation}
     \bar{I}=\frac{c^4}{G^2} \frac{I}{M^3},
    \label{eqn46}
    \end{equation}
where $I$ is $I^{(0)}$ for non-rotating configuration and $I_{tot}$ for the rotating configuration with c and G being the speed of light and gravitational constant, respectively. In all these figures rotation has its obvious effect in increasing moment of inertia.
   
   From physics point of view it is intuitive that, more is the compactness of the star less will be its deformability under external field and rotation. For this work, as already described in preceding paragraphs, secular instability does not occur. Hence, Fig.-\ref{fig12} \& Fig.-\ref{fig13}, the plots of rotational tidal love number and dimensionless rotational tidal love number with respect to the central density for different elements under different field strengths, show the nature of continuously increasing tidal deformability with decreasing central density. On the other hand, Fig.-\ref{fig14}, the plot of eccentricity, depicts the increasing nature of eccentricity while density is decreasing. Figs.-\ref{fig15}-\ref{fig17} are the plots showing the variations of quadrupole moment in unit of 10$^{49}$ gm.cm$^2$, normalized quadrupole moment and dimensionless quadrupole moment with respect to the central density for different elements under different field strengths. The normalization of quadrupole moment is done by dividing the spin quadrupolar moment by mass times the equatorial radius square of the non-rotating configuration. It is interesting to notice that, the tidal deformability, eccentricity and quadrupole moments, all these quantities follow almost an universal trend for all kind of white dwarfs except Iron white dwarfs under different field strengths. The Iron white dwarfs only show some deviations in all these cases. Due to presence of magnetic field, only the Iron white dwarfs show significant changes in the mass-radius relationship. In presence of magnetic field Iron white dwarfs consist of huge cores compared to other white dwarfs as the magnetic field energy and pressure contributions have been taken into the equation of state explicitly \cite{Roy2019}. Then, it turns out that, the eccentricity in the higher central density region increases for Iron white dwarfs [Fig.-\ref{fig14}]. As, lower is the eccentricity higher is the universality in the emergent properties of the compact stars, the Iron white dwarfs deviate most \cite{Yagi2014, Yip2017, Sham2015}. It is also worth mentioning that, higher the central density more is the universality in the trends. 

   Fig.-\ref{fig18} is the plot between the first pair of the parameters (I-Love-Q) which follow the universality under certain circumstances i.e. the dimensionless moment of inertia is plotted against the dimensionless love number. The definition of dimensionless moment of inertia is given in Eq.-(\ref{eqn46}) and the dimensionless love number can be defined as follows,
   
   \begin{equation}
    \bar{\lambda}=\frac{c^{10}}{G^4} \frac{\lambda}{M^5}.
   \label{eqn47}
   \end{equation}
All the calculated data points in this figure can well be fitted to a straight line given by, $\ln(\bar{I}) = -0.800188 + 0.412597 ~ \ln(\bar{\lambda})$ and hence is obeying an universal relationship. It is the case of the Iron again for which, the relative error is highest $\sim$ 1.5\% in the higher central density region for the highest field strength. As the field strength is reduced and the central density is decreased the universality is better followed like He, C, O white dwarfs.

   In Fig.-\ref{fig19}, dimensionless moment of inertia has been plotted with respect to the dimensionless quadrupole moment. Though all the points go well through a single straight line, $\ln(\bar{I}) = -4.07879 + 2.11982~\ln(\bar{Q})$, it turns out that, under high magnetic field and with higher central densities Iron white dwarfs tend to show some deviation from this trend. The maximum deviation shown by Iron is $\sim$3.5\% which again reduces with the decrease in the central density and field strength. The dimensionless quadrupole moment used, is defined as,

   \begin{equation}
    \bar{Q}=\frac{c^2Q}{I^{(0)}\Omega^2} M^{(0)},
   \label{eqn48}
   \end{equation}
where quadrupole moment Q can be found from Eq.-(\ref{eqn36}) and $\Omega$ is the angular velocity of the white dwarf.

   The dimensionless quadrupole moment and the dimensionless love number relationship has been depicted in Fig.-\ref{fig20} and the relationship can again be presented by a single straight line given by, $\ln(\bar{Q})=1.55013 + 0.194545 ~\ln(\bar{\lambda})$. Iron white dwarfs show again the maximum deviation from the universality in the higher central density region under higher field strengths. Fig.-\ref{fig21} shows the universality in the relationship between the normalized moment of inertia and the eccentricity. Similarly, normalized moment of inertia is plotted in Fig.-\ref{fig22} with respect to the normalized spin quadrupole moment of the magnetized white dwarfs under different magnetic field strengths and all the calculated points for different configurations can be fitted through a single curve and the universality turns out to be better than that in Fig.-\ref{fig19}. Finally, normalized spin quadrupole moment of the magnetized white dwarfs has been plotted against the eccentricity in Fig.-\ref{fig23} to check the universality relationship between these parameters. As the Figs.-\ref{fig21}-\ref{fig23}, depicting I-eccentricity-Q relationships, show lower relative errors compared to Figs.-\ref{fig18}-\ref{fig20}, it can be inferred that, I-eccentricity-Q relationship is better universal relationship than I-Love-Q relationship.

\end{document}